# Metamaterial Superconductors

Igor I. Smolyaninov [1,2)], Vera N. Smolyaninova [3)]

[1] *Department of Electrical and Computer Engineering, University of Maryland, College Park, MD 20742, USA ;  smoly@umd.edu*

[2] *Saltenna LLC, 1751 Pinnacle Drive #600 McLean, VA 22102,USA*

[3] *Department of Physics Astronomy and Geosciences, Towson University, 8000 York Rd., Towson, MD 21252, USA*

Abstract:

**Searching for natural materials exhibiting larger electron-electron interactions constitutes a traditional approach to high temperature superconductivity research. Very recently we pointed out that the newly developed field of electromagnetic metamaterials deals with the somewhat related task of dielectric response engineering on a sub-100 nm scale. Considerable enhancement of the electron-electron interaction may be expected in such metamaterial scenarios as in epsilon near zero (ENZ) and hyperbolic metamaterials. In both cases dielectric function may become small and negative in substantial portions of the relevant four-momentum space, leading to enhancement of the electron pairing interaction. This approach has been verified in experiments with aluminium-based metamaterials. Metamaterial superconductor with Tc = 3.9 K have been fabricated, that is three times that of pure aluminium (Tc = 1.2 K), which opens up new possibilities to considerably improve Tc of other simple superconductors. Taking advantage of the demonstrated success of this approach, the critical temperature of hypothetic niobium, MgB2 and $H_2S$-based metamaterial superconductors is evaluated. The $MgB_2$-based metamaterial superconductors are projected to reach the liquid nitrogen temperature range. In the case of an $H_2S$-based metamaterial projected Tc appears to reach ~250 K.**

**Key words:**   superconductivity; electromagnetic metamaterials; dielectric properties; epsilon near zero metamaterial; hyperbolic metamaterial

**Table of content**                                                    page

## 1. Introduction

One of the most important goals of condensed matter physics is reliably designing new materials with enhanced superconducting properties. Recently, an electromagnetic metamaterial strategy, consisting of deliberately engineering the dielectric properties of a nanostructured "metamaterial superconductor" that results in an enhanced electron pairing interaction that increases the value of the superconducting energy gap and the critical temperature, $T_c$, was suggested to achieve this goal [1, 2]. Our recent experimental work [3-5] has conclusively demonstrated that this approach can indeed be used to increase the critical temperature of a composite superconductor-dielectric metamaterial. For example, we have demonstrated the use of $Al_2O_3$-coated aluminum nanoparticles to form epsilon near zero (ENZ) core-shell metamaterial superconductors with a $T_c$ that is three times that of pure aluminum [4]. This deep and non-trivial connection between the fields of electromagnetic metamaterials and superconductivity research stems from the fact that superconducting properties of a material, such as electron-electron pairing interaction, the superconducting critical temperature $T_c$, etc. may be expressed via the effective dielectric response function $\varepsilon_{eff}(q, \omega)$ of the material [6]. For example, considerable enhancement of attractive electron-electron interaction may be expected in such actively studied metamaterial scenarios as epsilon near zero (ENZ) [7] and hyperbolic metamaterials [8] since in both cases $\varepsilon_{eff}(q, \omega)$ may become small and negative in substantial portions or the four-momentum $(q, \omega)$ space. Such an effective dielectric response-based macroscopic electrodynamics description is valid if the material may be considered as a homogeneous medium on the spatial scales below the superconducting coherence length.

The metamaterial superconductor approach takes advantage of the recent progress in plasmonics [9] and electromagnetic metamaterials [10] to engineer an artificial medium or "metamaterial", so that its effective dielectric response function $\varepsilon_{eff}(q,\omega)$ conforms to almost any desired behaviour. It appears natural to use this newly found freedom to engineer and maximize the electron pairing interaction in such an artificial superconductor via engineering its dielectric response function $\varepsilon_{eff}(q,\omega)$. In this article we will review the comparative advantages and shortcomings of various electromagnetic metamaterial strategies to achieve this goal. For example, the ENZ approach based on the plasmonic core-shell nanostructures [4] seems to be the most efficient way to achieve the highest $T_c$ values. However, the core-shell metamaterial superconductor geometry exhibits poorer transport properties compared to its parent (aluminum) superconductor. A natural way to overcome this issue is the implementation of the layered hyperbolic metamaterial geometry, which is based on parallel periodic layers of metal separated by layers of dielectric. This geometry ensures excellent transport properties in the plane of the layers. As noted in [11], typical high $T_c$ superconductors (such as BSCCO) exhibit hyperbolic behaviour in a substantial portion of the far infrared and THz frequency ranges. Indeed, it was demonstrated that the artificial hyperbolic metamaterial geometry may also lead to a considerable enhancement of superconducting properties [5].

We should also mention that very recently other approaches to artificial metamaterial superconductor engineering have been proposed by several groups, such as suggestions to create better superconductors by phonon spectrum modification via periodic nanostructuring of thin film [12] and bulk [13] superconductors. It was also proposed to modify natural hyperbolic superconductors (such as $HgBa_2CuO_{4+y}$) via inhomogeneous charge density waves (CDW) [14]. Together with our electromagnetic metamaterial approach these strategies open up new areas in superconductivity research at the unconventional intersection of nanophotonics and solid state physics, which may potentially reach the scientific dream of achieving room temperature superconductivity [15].

## 2. Macroscopic electrodynamic description of superconductivity

Electromagnetic properties are known to play a very important role in the pairing mechanism of superconductors [6]. According to the BCS theory, a Cooper pair is formed from two electrons with opposite spins and momenta that are loosely bound. This mechanism may be described as an attractive interaction of electrons that results from the polarization of the ionic lattice which these electrons create as they move through the lattice. Based on this interpretation, Kirzhnits *et al.* formulated their description of superconductivity in terms of the dielectric response function of the superconductor [6]. They demonstrated that the electron-electron interaction in a superconductor may be expressed in the form of an effective Coulomb potential

$$V(\vec{q},\omega) = \frac{4\pi e^2}{q^2 \varepsilon_{eff}(\vec{q},\omega)} = V_C \frac{1}{\varepsilon_{eff}(\vec{q},\omega)} \qquad (1)$$

where $V_C$ is the Fourier-transformed Coulomb potential in vacuum, and $\varepsilon_{eff}(q,\omega)$ is the linear dielectric response function of the superconductor treated as an effective medium. While the thermodynamic stability condition implies that $\varepsilon_{eff}(0,0)>0$, the dielectric response function at higher frequencies and momenta may become large and negative, which accounts for the weak net attraction and pairing of electrons in the superconducting condensate. Based on this approach, Kirzhnits et al. derived explicit expressions for the superconducting gap $\Delta$ and critical temperature, $T_c$, of the superconducting transition as a function of the dielectric response function, and demonstrated that their expressions agree with the BCS result.

The key observation made in Refs. [1,2] was that the "homogeneous medium" approximation may remain valid even if the basic structural elements of the material are not simple atoms or molecules. It was suggested that artificial "metamaterials" may be created from much bigger building blocks, and the electromagnetic properties of these fundamental building blocks may be engineered in such a way that the attractive pairing interaction described by Eq. (1) is maximized. The two potential solutions described in [1,2] require minimization of $\varepsilon_{eff}(q,\omega)$ within a substantial portion of the relevant four-momentum spectrum ($|\vec{q}| \leq 2k_F$, $\omega \leq$ BCS cutoff around the Debye energy), while keeping the real part of $\varepsilon_{eff}(q,\omega)$ negative. These solutions involve either the epsilon-near-zero (ENZ) [7], or hyperbolic metamaterial [8] scenarios, which naturally lead to broadband small and negative $\varepsilon_{eff}(q,\omega)$ behavior in substantial portions of the relevant four-momentum spectrum. In the isotropic ENZ scenario the effective pairing interaction is described by Eq. (1), where $\varepsilon_{eff}(q,\omega)$ is assumed to characterize the macroscopic electrodynamic properties of the composite metamaterial. On the other hand, in the hyperbolic metamaterial scenario the effective Coulomb potential from Eq. (1) assumes the form

$$V(\vec{q},\omega) = \frac{4\pi e^2}{q_z^2 \varepsilon_2(\vec{q},\omega) + (q_x^2 + q_y^2)\varepsilon_1(\vec{q},\omega)} \qquad (2)$$

where $\varepsilon_{xx} = \varepsilon_{yy} = \varepsilon_1$ and $\varepsilon_{zz} = \varepsilon_2$ have opposite signs. As a result, the effective Coulomb interaction of two electrons may become attractive and very strong along spatial directions where

$$q_z^2 \varepsilon_2(\vec{q},\omega) + (q_x^2 + q_y^2)\varepsilon_1(\vec{q},\omega) \approx 0 \quad , \qquad (3)$$

which indicates that the superconducting order parameter must be strongly anisotropic. This situation resembles the hyperbolic character of such high $T_c$ superconductors as BSCCO [11]. However, engineering such ENZ or hyperbolic metamaterials constitutes a much more challenging task compared to typical applications of superconducting metamaterials suggested so far [16], which only deal with metamaterial engineering on scales which are much smaller than the microwave or RF wavelengths. Since the superconducting coherence length (the size of the Cooper pair) is $\xi\sim100$ nm in a typical BCS superconductor [17], the metamaterial unit size must fall within the window between ~0.3 nm (given by the atomic scale) and $\xi\sim100$ nm scale of a typical Cooper pair. Moreover, the coherence length of the metamaterial superconductor must be determined in a self-consistent manner. The coherence length will decrease with increasing $T_c$ of the metamaterial superconductor since the approach of Kirzhnits et al. gives rise to the same BCS-like relationship between the superconducting gap $\Delta$ and the coherence length $\xi$ [17]:

$$\left(\frac{\xi}{V_F}\right)\Delta \sim \hbar, \qquad (4)$$

where $V_F$ is the Fermi velocity. Therefore, the metamaterial structural parameter (such as the inter-layer or inter-particle distance, etc.) which must remain smaller than the coherence length, defines the limit of critical temperature enhancement. Demonstration of the superconductivity enhancement in electromagnetic metamaterials opens up numerous new possibilities for metamaterial enhancement of $T_c$ in such practically important simple superconductors as niobium and $MgB_2$.

## 3. Epsilon near zero superconductors: theory and experiment

Let us demonstrate that tuning electron-electron interaction is indeed possible using the epsilon-near-zero (ENZ) metamaterial approach, which is based on intermixing metal and dielectric components in the right proportions. Following [18], a simplified dielectric response function of a metal may be written as

$$\varepsilon_m\left(q,\omega\right) = \left(1 - \frac{\omega_p^2}{\omega^2 + i\omega\Gamma - \omega_p^2 q^2 / k^2}\right)\left(1 - \frac{\Omega_1^2(q)}{\omega^2 + i\omega\Gamma_1}\right)...\left(1 - \frac{\Omega_n^2(q)}{\omega^2 + i\omega\Gamma_n}\right) \tag{5}$$

where $\omega_p$ is the plasma frequency, $k$ is the inverse Thomas-Fermi radius, $\Omega_n(q)$ are dispersion laws of various phonon modes, and $\Gamma_n$ are the corresponding damping rates. Zeroes of the dielectric response function of the bulk metal (which correspond to various bosonic modes) maximize electron-electron pairing interaction given by Eq. (1). In the discussion below we will limit our consideration to the behaviour of $\varepsilon_m(q,\omega)$ near $\omega = \Omega_1(q)$ in the vicinity of the Fermi surface, so that a more complicated representation of $\varepsilon_m(q,\omega)$ by the Lindhard function, which accurately represents $\varepsilon_m(q,\omega)$ in the $q \to 0$ and $\omega \to 0$ limit [18] does not need to be used. We are going to use only the fact that $\varepsilon_m$ changes sign near $\omega = \Omega_1(q)$ by passing through zero. Expression for $\varepsilon_m(q,\omega)$ given by Eq.(5) does exhibit this behavior.

As summarized in [19], the critical temperature of a superconductor in the weak coupling limit is typically calculated as

$$T_c = \theta \, \exp\left(-\frac{1}{\lambda_{eff}}\right), \tag{6}$$

where $\theta$ is the characteristic temperature for a bosonic mode mediating electron pairing (such as the Debye temperature $\theta_D$ in the standard BCS theory, or $\theta_{ex} = \hbar\omega_{ex}/k$ or $\theta_{pl} = \hbar\omega_{pl}/k$ in the theoretical proposals which suggest excitonic or plasmonic mediation of the electron pairing interaction), and $\lambda_{eff}$ is the dimensionless coupling constant defined by $V(q,\omega) = V_C(q)\,\varepsilon^{-1}(q,\omega)$ and the density of states $\nu$ (see for example [20]):

$$\lambda_{eff} = -\frac{2}{\pi}\nu\int\limits_0^\infty \frac{d\omega}{\omega}\left\langle V_C(q)\,\mathrm{Im}\,\varepsilon^{-1}(\vec{q},\omega)\right\rangle,\tag{7}$$

where $V_C$ is the unscreened Coulomb repulsion, and the angle brackets denote average over the Fermi surface.

Compared to the bulk metal, zeroes of the effective dielectric response function $\varepsilon_{eff}(q,\omega)$ of the metal-dielectric metamaterial are observed at shifted positions compared to the zeroes of $\varepsilon_m(q,\omega)$ [2], and additional zeros may also appear. According to the Maxwell-Garnett approximation [21], mixing of nanoparticles of a superconducting "matrix" with dielectric "inclusions" (described by the dielectric constants $\varepsilon_m$ and $\varepsilon_d$, respectively) results in the effective medium with a dielectric constant $\varepsilon_{eff}$, which may be obtained as

$$\left(\frac{\varepsilon_{eff} - \varepsilon_m}{\varepsilon_{eff} + 2\varepsilon_m}\right) = (1-n)\left(\frac{\varepsilon_d - \varepsilon_m}{\varepsilon_d + 2\varepsilon_m}\right),\tag{8}$$

where $n$ is the volume fraction of metal ($0 \le n \le 1$). The explicit expression for $\varepsilon_{eff}$ may be written as

$$\varepsilon_{eff} = \frac{\varepsilon_m\left((3-2n)\varepsilon_d + 2n\varepsilon_m\right)}{\left(n\varepsilon_d + (3-n)\varepsilon_m\right)},\tag{9}$$

and it is easy to verify that

$$\varepsilon_{eff}^{-1} = \frac{n}{(3-2n)}\frac{1}{\varepsilon_m} + \frac{9(1-n)}{2n(3-2n)}\frac{1}{(\varepsilon_m + (3-2n)\varepsilon_d / 2n)} \qquad (10)$$

For a given value of the metal volume fraction $n$, the ENZ conditions ($\varepsilon_{eff} \approx 0$) may be obtained around $\varepsilon_m \approx 0$ and around

$$\varepsilon_m \approx -\frac{3-2n}{2n}\varepsilon_d \qquad (11)$$

Since the absolute value of $\varepsilon_m$ is limited by some value $\varepsilon_{m,max}$ (see Eq. (5)), the latter zero of $\varepsilon_{eff}$ disappears at small $n$ as $n \to 0$ at some critical value of the volume fraction $n = n_{cr} = 1.5\varepsilon_d/\varepsilon_{m,max}$. Let us evaluate $\mathrm{Im}(\varepsilon^{-1}_{eff}(q,\omega))$ near this zero of $\varepsilon_{eff}$ at $n > n_{cr}$. Based on Eqs. (10),

$$\mathrm{Im}\left(\varepsilon^{-1}_{eff}\right) \approx \frac{9(1-n)}{2n(3-2n)\varepsilon''_m} \qquad (12)$$

where $\varepsilon_m$'$= \mathrm{Re}\varepsilon_m$ and $\varepsilon_m$''$= \mathrm{Im}\varepsilon_m$. Assuming that $\nu \sim n$ (which is justified by the fact that there are no free charges in the dielectric phase of the metamaterial) and using Eq. (7), we may obtain the resulting expression for $\lambda_{eff}$ as a function of $\lambda_m$ and $n$:

$$\lambda_{eff} \approx \frac{9(1-n)}{2(3-2n)}\lambda_m, \qquad (13)$$

where $\lambda_m$ is also determined by Eq.(7) in the limit $\varepsilon \to \varepsilon_m$, and we have assumed the same magnitudes of $\varepsilon_m$'' for the metamaterial zero described by Eq. (11) and for the $\varepsilon_m \approx 0$ conditions. This assumption will be re-evaluated in Section 5 and confronted with experimental data for aluminium-based ENZ metamaterial, which indicate that $\varepsilon_m$'' changes by ~11%. Fig. 1a shows the predicted behaviour of $\lambda_{eff}/\lambda_m$ as a function of $n$. Based on this prediction, we may expect enhancement of superconducting properties of the metal-dielectric metamaterial in the $n_{cr} < n < 0.6$ range. For comparison, Fig. 1a also shows the behaviour of $\lambda_{\text{eff}}$ near the $\varepsilon_m \approx 0$ pole of the inverse dielectric response function of the metamaterial. According to Eq. (9), near $\varepsilon_m \approx 0$ the effective dielectric response function of the metamaterial behaves as

$$\varepsilon_{\text{eff}} \approx \frac{3-2n}{n} \varepsilon_m \qquad (14)$$

Therefore, near this pole

$$\lambda_{\text{eff}} \approx \frac{n^2}{(3-2n)} \lambda_m \qquad (15)$$

Let us apply the obtained simple estimates to the case of Al-Al$_2$O$_3$ core-shell metamaterial studied in [4]. Assuming the known values $Tc_{\text{bulk}} = 1.2$ K and $\theta_D = 428$ K of bulk aluminium [8], Eq. (6) results in $\lambda_m = 0.17$, which corresponds to the weak coupling limit. In order to make the mechanism behind the enhancement of $T_c$ in the Al-Al$_2$O$_3$ core-shell metamaterial superconductor abundantly clear, we have plotted the hypothetic values of $T_c$ as a function of metal volume fraction $n$, which would originate from either Eq. (13) or Eq. (15) in the absence of each other. The corresponding values calculated as

$$T_c = T_{Cbulk} \exp\left( \frac{1}{\lambda_m} - \frac{1}{\lambda_{\text{eff}}} \right) \qquad (16)$$

are shown in Fig. 1b. The vertical dashed line corresponds to the assumed value of $n_{\text{cr}}$. The experimentally measured data points from Ref. 4 are shown for comparison on the same plot. The match between the experimentally measured values of enhanced $T_c$ and the theoretical curve obtained based on Eq. (13) is quite impressive, given the fact that Eqs. (13) and (16) do not contain any free parameters. Such a good match unambiguously identifies metamaterial enhancement as the physical mechanism of critical temperature tripling in the Al-Al$_2$O$_3$ core-shell metamaterial superconductor. It is impressive that such simple and well known nanophotonics tools as the effective medium theory may be used to considerably enhance the critical temperature of simple superconductors.

The proof of principle experiments conducted with both random [3] and core-shell [4] ENZ metal-dielectric metamaterials indeed validate the effective medium-based theory outlined above. Our initial superconducting metamaterial samples [3] were prepared using commercially available tin and barium titanate nanoparticles obtained from the US Research Nanomaterials, Inc. The nominal diameter of the BaTiO$_3$ nanoparticles was 50 nm, while tin nanoparticle size was specified as 60-80 nm. Both nanoparticle sizes fall substantially below the superconducting coherence length in pure tin $\xi_{Sn}$ ~230 nm [17]. Our choice of materials was based on results of numerical calculations of the real part of the dielectric constant of the Sn/BaTiO$_3$ mixture shown in Fig.2. These calculations were based on the measured dielectric properties of Sn [23] and BaTiO$_3$ [24], respectively, and on the Maxwell-Garnett expression, Eq. (9), for the dielectric permittivity of the mixture. These calculations indicated that it was possible to achieve broadband ENZ conditions in the 30-50% range of the volume fraction of BaTiO$_3$ in the frequency range of relevance around $h\nu \sim kT_c$ of pure tin.

The Sn/BaTiO$_3$ superconducting metamaterials were fabricated by combining the given amounts of Sn and BaTiO$_3$ nanoparticle powders by volume into a single test tube filled with de-ionized water. The resulting suspensions were sonicated and magnetically stirred for 30 minutes, followed by water evaporation. The remaining mixtures of Sn and BaTiO$_3$ nanoparticles were compressed into test pellets using a hydraulic press. A typical Scanning Electron Microscopy (SEM) image of the resulting

Sn/BaTiO$_3$ composite metamaterial is shown in Fig. 3. The original compressed nanoparticles are clearly visible in the image. The resulting average metamaterial composition has been verified after fabrication using an SEM with energy dispersive X-ray spectroscopy (EDS). The compositional analysis of an area of the samples of about 1 μm in diameter is consistent with this nominal composition. Such EDX spectra were used to establish the homogeneous character of the fabricated composite metamaterials.

The superconducting critical temperature $T_c$ of various Sn/BaTiO$_3$ metamaterials was measured via the onset of diamagnetism for samples with different volume fractions of BaTiO$_3$ using a MPMS SQUID magnetometer. The zero field cooled (ZFC) magnetization per unit mass for several samples with varying concentrations of BaTiO$_3$ is plotted in Fig. 4(a). The temperatures of the onset of the superconducting transition and the temperatures of the midpoint of the transition are plotted in Fig. 5. The $T_c$ increased from the pure Sn value of 3.68 K with increasing BaTiO$_3$ concentration to a maximum $\Delta T_c \sim 0.15$ K or 4% compared to the pure tin sample for the 40% sample followed by $T_c$ decreasing at higher volume fractions. A pure tin sample was prepared from pressed tin nanoparticles of the same diameter using the same method of preparation. The value of $T_c$ agreed with expected value of pure Sn. The magnetization of the composite samples in the superconducting state is comparable in magnitude with the pure tin sample. The increase of $T_c$ and its dependence on effective dielectric constant determined by the concentration of BaTiO$_3$ agrees qualitatively with the Maxwell-Garnett theory-based calculations described above (see Fig. 2). Moreover, decrease of $T_c$ at higher volume fraction of BaTiO$_3$, which is apparent from Fig. 5 also agrees well with the Maxwell-Garnett theory, since $\varepsilon_{eff}(q, \omega)$ changes sign to positive for higher BaTiO$_3$ concentrations.

In order to verify the reproducibility of these results, our measurements were repeated for several sets of nanocomposite samples fabricated using strontium titanate nanoparticles (instead of barium titanate) using the same fabrication technique described above. The superconducting critical temperature $T_c$ of the Sn/SrTiO$_3$ metamaterial samples was measured for samples with different volume fractions of SrTiO$_3$ nanoparticles as shown in Fig. 4(b) producing volume fraction dependencies, which are similar to those obtained for Sn/BaTiO$_3$ samples. These results established that the metamaterial approach to dielectric response engineering can indeed increase the critical temperature of a composite superconductor-dielectric metamaterial. However, despite this initial success, the observed critical temperature increase was modest. It was argued in [2] that the random nanoparticle mixture geometry may not be ideal because simple mixing of superconductor and dielectric nanoparticles results in substantial spatial variations of $\varepsilon_{eff}(q,\omega)$ throughout a metamaterial sample. Such variations lead to considerable broadening and suppression of the superconducting transition.

To overcome this issue, it was suggested that an ENZ plasmonic core-shell metamaterial geometry, which has been developed to achieve partial cloaking of macroscopic objects [25], should be implemented [2]. The cloaking effect relies on mutual cancellation of scattering by the dielectric core (having $\varepsilon_d > 0$) and plasmonic shell (with $\varepsilon_m < 0$) of the nanoparticle, so that the effective dielectric constant of the nanoparticle becomes very small and close to that of vacuum (a plasmonic core with a dielectric shell may also be used). This approach may be naturally extended to the core-shell nanoparticles having negative ENZ behaviour, as required in the superconducting application. Synthesis of such individual ENZ core-shell nanostructures followed by nanoparticle self-assembly into a bulk ENZ metamaterial (as shown in Fig.6) appears to be a viable way to fabricate an extremely homogeneous metamaterial superconductor.

The design of an individual core-shell nanoparticle is based on the fact that scattering of an electromagnetic field by a sub-wavelength object is dominated by its electric dipolar contribution, which is defined by the integral sum of its volume polarization [25]. A material with $\varepsilon > 1$ has a positive electric polarizability, while a material with $\varepsilon < 1$ has a negative electric polarizability (since the local electric polarization vector, $P = (\varepsilon - 1)E/4\pi$, is opposite to $E$). As a result, the presence of a plasmonic shell (core) cancels the scattering produced by the dielectric core (shell), thus providing a cloaking effect. Similar consideration for the negative ENZ case leads to the following condition for the core-shell geometry:

$$r_c^3 \varepsilon_c \approx -\left(r_s^3 - r_c^3\right)\varepsilon_s ,\tag{17}$$

where $r_c$ and $r_s$ are the radii, and $\varepsilon_c$ and $\varepsilon_s$ are the dielectric permittivities of the core and shell, respectively. Eq.(17) corresponds to the average dielectric permittivity of the core-shell nanoparticle being approximately equal to zero. Working on the negative side of this equality will ensure negative ENZ character of each core-shell nanoparticle. A dense assembly of such core-shell nanoparticles will form a medium that will have small negative dielectric permittivity. Moreover, in addition to obvious advantage in homogeneity, a core-shell based metamaterial superconductor design enables tuning of the spatial dispersion of the effective dielectric permittivity $\varepsilon_{eff}(q, \omega)$ of the metamaterial, which would further enhance its $T_c$ [2]. Spatial dispersion of a metamaterial is indeed well known to originate from plasmonic effects in its metallic constituents. In a periodic core-shell nanoparticle-based ENZ metamaterial spatial dispersion is defined by the coupling of plasmonic modes of its individual nanoparticles. This coupling enables propagating plasmonic Bloch modes and, hence, nonlocal effects.

The successful realization of such an ENZ core-shell metamaterial superconductor using compressed $Al_2O_3$-coated aluminium nanoparticles has been reported in [4], leading to tripling of the metamaterial critical temperature compared to the bulk aluminium. This material is ideal for the proof of principle experiments because the critical temperature of aluminium is quite low ($Tc_{Al}$=1.2K [17]), leading to a very large superconducting coherence length $\xi$=1600 nm [17]. Such a large value of $\xi$ facilitates the metamaterial fabrication requirements while $Al_2O_3$ exhibits very large positive values of dielectric permittivity up to $\varepsilon_{Al2O3}$~200 in the THz frequency range [26]. These results provide an explanation for the long known, but not understood, enhancement of the $T_c$ of granular aluminum films [27,28].

The 18 nm diameter Al nanoparticles for these experiments were acquired from the US Research Nanomaterials, Inc. Upon exposure to the ambient conditions a ~ 2 nm thick $Al_2O_3$ shell is known to form on the aluminium nanoparticle surface [22], which is comparable to the 9 nm radius of the original Al nanoparticle. Further aluminium oxidation may also be achieved by heating the nanoparticles in air. The resulting core-shell $Al_2O_3$-Al nanoparticles were compressed into macroscopic, ~1cm diameter, ~0.5 mm thick test pellets using a hydraulic press, as illustrated in the inset in Fig.6.

The IR reflectivity of such core-shell metamaterial samples was measured in the long wavelength IR (LWIR) (2.5-22.5 μm) range using an FTIR spectrometer, and compared with reflectivity spectra of Al and $Al_2O_3$, as shown in Fig. 7. While the reflectivity spectrum of Al is almost flat, the spectrum of $Al_2O_3$ exhibits a very sharp step-like behaviour around 11 μm that is related to the phonon-polariton resonance, which results from coupling of an infrared photon with an optic phonon of $Al_2O_3$ [29]. The step in reflectivity is due to the negative sign of $\varepsilon_{Al2O3}$ near resonance. This step-like behaviour may be used to characterize the volume fraction of $Al_2O_3$ in the core-shell metamaterial. In the particular case shown in Fig.7, the volume fraction of $Al_2O_3$ in the core-shell metamaterial may be estimated, based on the Maxwell-Garnett approximation, as ~ 39%, which corresponds to $(r_s-r_c)$~$0.18r_c$. At $r_c$~9 nm the corresponding thickness of $Al_2O_3$ appears to be $(r_s-r_c)$~1.6 nm, which matches expectations based on [22].

The Kramers-Kronig analysis of the FTIR reflectivity spectra of the Al-$Al_2O_3$ sample also allows us to evaluate $\varepsilon_{eff}(0,\omega)$ for the metamaterial in the LWIR spectral range. Plots of the real part of $\varepsilon$ for pure Al and for the Al-$Al_2O_3$ core-shell metamaterial based on the Kramers-Kronig analysis of the data in Fig.7 are plotted in Fig.8. The plot in Fig.8(a) clearly demonstrates that $\varepsilon_{Al\text{-}Al2O3} \ll \varepsilon_{Al}$ so that the ENZ condition was achieved in the sense that the initial dielectric constant of aluminium was reduced by a factor ~1000. On the other hand, Fig.8(b) demonstrates that the dielectric constant of the Al-$Al_2O_3$ core-shell metamaterials remains negative and relatively small above 11 μm. In particular, the large negative contribution to $\varepsilon$ from the aluminium cores is compensated by the large positive contribution from the $Al_2O_3$ shells leading to the upturn of $\varepsilon_{Al\text{-}Al2O3}$ that is observed near 20 μm in Fig.8(b) which is caused by the large positive value of $\varepsilon_{Al2O3}$ in this spectral range. Note that while both metamaterials shown in Fig.8(b) exhibit much smaller $\varepsilon$ compared to the bulk aluminium, the metamaterial prepared using less oxidized aluminium nanoparticles exhibits considerably larger negative $\varepsilon$. The relatively large noise observed in the calculated plot of $\varepsilon_{Al}$ in Fig.8(a) is due to the fact that the aluminium reflectivity is close to 100% above 7 μm so that the Kramers-Kronig-based numerical analysis of the reflectivity data does not work reliably for pure aluminium samples in this spectral range. Another limitation on the accuracy of the analysis is the use of the finite spectral range (2.5-22.5 μm) of the FTIR spectrometer rather than the infinite one assumed by the rigorous Kramers-Kronig analysis. These limitations notwithstanding, we note that our result for pure aluminium is in good agreement with the tabulated data for $\varepsilon_{Al}$ reported in [30]. Therefore, these results reliably confirm the ENZ character of the core-shell Al-Al$_2$O$_3$ metamaterial. It is also interesting to note that the same FTIR technique applied to the tin-BaTiO$_3$ nanocomposite metamaterials studied in [3] also confirms their expected ENZ character as illustrated in Fig. 9. In both cases, the goal of metamaterial engineering was to create an effective superconducting medium with negative ENZ response. While both Figs. 8 and 9 confirm that this goal has been achieved, the core-shell geometry of the developed Al-Al$_2$O$_3$ metamaterial has a clear advantage. The core-shell geometry guarantees a homogeneous spatial distribution of the effective dielectric response function, leading to tripling of T$_c$ for the Al-Al$_2$O$_3$ core-shell metamaterial, compared to ~5% increase of T$_c$ of the tin-BaTiO$_3$ random nanocomposite metamaterial developed in [3].

The T$_c$ of various Al-Al$_2$O$_3$ core-shell metamaterials was determined via the onset of diamagnetism for samples with different degrees of oxidation using a MPMS SQUID magnetometer. The zero field cooled (ZFC) magnetization per unit mass versus temperature for several samples with various volume fractions of Al$_2$O$_3$ is plotted in Fig. 10(a), while the corresponding reflectivity data are shown in Fig.10(b). Even though the lowest achievable temperature with our MPMS SQUID magnetometer was 1.7K, we were able to observe a gradual increase of T$_c$ that correlated with an increase of the Al$_2$O$_3$ volume fraction as determined by the drop in reflectivity shown in Fig.10(b). The reliability of the MPMS SQUID magnetometer at T>1.7K was checked by measurements of the $T_c$ of bulk tin at $T_c=3.7K$, as described in ref.[3], in excellent agreement with the textbook data [17].

The observed increase in $T_c$ also showed good correlation with the results of the Kramers-Kronig analysis shown in Fig.8(b): samples exhibiting smaller negative $\varepsilon$ demonstrated higher $T_c$ increase. The highest onset temperature of the superconducting transition reached 3.9K, which is more than three times as high as the critical temperature of bulk aluminium, $T_{cAl}=1.2K$ [17]. All of the samples exhibited a small positive susceptibility that increased with decreasing temperature, consistent with the presence of small amounts of paramagnetic impurities. The discussed $T_c$ values were determined by the beginning of the downturn of M(T), where the diamagnetic superconducting contribution starts to overcome paramagnetic contribution, making this temperature the lower limit of the onset of superconductivity. Further oxidation of aluminium nanoparticles by annealing for 2 hours at 600°C resulted in a $T_c$ less than 1.7K, our lowest achievable temperature. Based on the reflectivity step near 11 μm (see Fig.10(b)), the volume fraction of $Al_2O_3$ in this sample may be estimated as ~ 50%, which corresponds to $(r_s-r_c)\sim0.26r_c$. For $r_c\sim9$ nm, the corresponding thickness of $Al_2O_3$ was $(r_s-r_c)\sim2.4$ nm.

Thus, the theoretical prediction of a large increase of $T_c$ in ENZ core-shell metamaterials shown in Fig.1 has been confirmed by direct measurements of $\varepsilon_{eff}(0,\omega)$ of the fabricated metamaterials and the corresponding measurements of the increase of $T_c$. These results strongly suggest that increased aluminium $T_c$'s that were previously observed in very thin (<50 nm thickness) granular aluminium films [27,28] and disappeared at larger film thicknesses were due to changes in the dielectric response function rather than quantum size effects and soft surface phonon modes [28]. As clearly demonstrated by the experimental data and the discussion above, the individual Al nanoparticle size is practically unaffected by oxidation, thus excluding the size effects as an explanation of giant $T_c$ increase in our bulk core-shell metamaterial samples. The developed technology enables efficient nanofabrication of bulk aluminium-based metamaterial superconductors with a $T_c$ that is three times that of pure aluminium and with virtually unlimited shapes and dimensions. These results open up numerous new possibilities of considerable $T_c$ increase in other simple superconductors.

## 4. Hyperbolic metamaterial superconductors: theory and experiment

While the Tc increase observed in ENZ metamaterial superconductors is impressive, the core-shell metamaterial superconductor geometry exhibits poor transport properties compared to its parent (aluminum) superconductor. A natural way to overcome this issue is the implementation of the hyperbolic metamaterial geometry (Fig. 11a), which has been suggested in [1, 2]. Hyperbolic metamaterials are extremely anisotropic uniaxial materials, which behave like a metal ($\mathrm{Re}\varepsilon_{xx} = \mathrm{Re}\varepsilon_{yy} < 0$) in one direction and like a dielectric ($\mathrm{Re}\varepsilon_{zz} > 0$) in the orthogonal direction. Originally introduced to overcome the diffraction limit of optical imaging [8], hyperbolic metamaterials demonstrate a number of novel phenomena resulting from the broadband singular behaviour of their density of photonic states. The "layered" hyperbolic metamaterial geometry shown in Fig. 11a is based on parallel periodic layers of metal separated by layers of dielectric. This geometry ensures excellent transport properties in the plane of the layers. Let us demonstrate that the artificial hyperbolic metamaterial geometry may also lead to a considerable enhancement of superconducting properties.

The first successful realization of an artificial hyperbolic metamaterial superconductor has been reported in [5]. The metamaterial was made of aluminum films separated by thin layers of $Al_2O_3$. This combination of materials is ideal for the proof of principle experiments because it is easy to controllably grow $Al_2O_3$ on the surface of Al, and because the critical temperature of aluminum is quite low ($T_{c\ Al} = 1.2$ K [17]), leading to a very large superconducting coherence length $\xi = 1600$ nm [17]. Such a large value of $\xi$ facilitates the metamaterial fabrication requirements since the validity of macroscopic electrodynamic description of the metamaterial superconductor requires that its spatial structural parameters must be much smaller than $\xi$. It appears that the $Al/Al_2O_3$ hyperbolic metamaterial geometry is capable of superconductivity enhancement, which is similar to that observed for a core-shell metamaterial geometry [4], while having much better transport and magnetic properties compared to the core-shell superconductors. The multilayer $Al/Al_2O_3$ hyperbolic metamaterial samples for these experiments (Fig. 12) were prepared using sequential thermal evaporation of thin aluminum films followed by oxidation of the top layer for 1 hour in air at room temperature. The first layer of aluminum was evaporated onto a glass slide surface. Upon exposure to ambient conditions a ~ 2 nm thick $Al_2O_3$ layer is known to form on the aluminum film surface [22]. Further aluminum oxidation may also be achieved by heating the sample in air. The oxidized aluminum film surface was used as a substrate for the next aluminum layer. This iterative process was used to fabricate thick multilayer (up to 16 metamaterial layers) $Al/Al_2O_3$ hyperbolic metamaterial samples. A transmission electron microscope (TEM) image of the multilayer metamaterial sample is shown in Fig. 12. During TEM experiments samples were coated with gold and platinum to ensure conductivity of the surface during sample preparation. A focused ion beam (FIB) microscope was used to prepare a sample for transmission electron microscopy (TEM). Samples were analyzed using a JEOL 2200 FS TEM, acquiring bright field and high resolution images. Fig. 12 shows an image from the hyperbolic stack, showing polycrystalline Al grains, with 1-2 nm thick amorphous $Al_2O_3$ spacing layers, corresponding with the designed structure. Some $Al_2O_3$ layers are difficult to discern due to slight sample warping from the preparation process. The inset shows that the interfacial layers are indeed amorphous, between polycrystalline grains of Al, that, in cross-section, exhibit Moire cross-hatching.

To demonstrate that these multilayer samples exhibit hyperbolic behaviour, we studied their transport and optical properties (Figs. 13-15). The temperature dependences of the sheet resistance of a 16-layer 10nm/layer $Al/Al_2O_3$ hyperbolic metamaterial and a 100 nm thick Al film are shown in Fig. 13a. As illustrated in the logarithmic plot in the inset, the electronic conductivity of the metamaterial approaches conductivity values of bulk aluminum (indicated by the arrow), and is far removed from the parameter space characteristic for granular Al films [28], which is indicated by the gray area in the inset. These results were corroborated by measurements of IR reflectivity of these samples, shown in Fig. 13b. The IR reflectivity of the hyperbolic metamaterial samples was measured in the long wavelength IR (LWIR) (2.5-22.5 µm) range using an FTIR spectrometer, and compared with reflectivity spectra of Al and $Al_2O_3$. While the reflectivity spectrum of bulk Al is almost flat, the spectrum of $Al_2O_3$ exhibits a very sharp step-like behavior around 11 µm that is related to the phonon-polariton resonance, which results from coupling of an infrared photon with an optical phonon of $Al_2O_3$ [29]. The step in reflectivity is due to the negative sign of $\varepsilon_{Al2O3}$ near the resonance. The absence of this step in our multilayers indicates that the aluminum layers in these samples are continuous and not intermixed with aluminum oxide. Such a step is clearly observed in reflectivity data obtained from a core-shell Al/Al$_2$O$_3$ sample shown in Fig. 13b where the aluminum grains are separated from each other by Al$_2$O$_3$. On the other hand, this step is completely missing in reflectivity spectra of the hyperbolic metamaterial samples (the step at 9 μm observed in the spectrum of a three-layer sample is due to phonon-polariton resonance of the SiO$_2$ substrate). Thus, the transport and optical measurements confirm excellent DC and AC (optical) conductivity of the aluminum layers of the fabricated hyperbolic metamaterials.

The Kramers-Kronig analysis of the FTIR reflectivity spectra of Al and Al$_2$O$_3$ measured in [4] also allowed us to calculate the $\varepsilon_1$ and $\varepsilon_2$ components of the Al/Al$_2$O$_3$ layered films in the LWIR spectral range using the Maxwell-Garnett approximation as follows:

$$\varepsilon_1 = n\varepsilon_m + (1-n)\varepsilon_d \tag{18}$$

$$\varepsilon_2 = \frac{\varepsilon_m \varepsilon_d}{(1-n)\varepsilon_m + n\varepsilon_d} \tag{19}$$

where $n$ is the volume fraction of metal, and $\varepsilon_m$ and $\varepsilon_d$ are the dielectric permittivities of the metal and dielectric, respectively [31]. Results of these calculations for a multilayer metamaterial consisting of 13 nm thick Al layers separated by 2 nm of Al$_2$O$_3$ are shown in Fig. 14. The metamaterial appears to be hyperbolic except for a narrow LWIR spectral band between 11 and 18 μm. A good match between the Maxwell-Garnett approximation (Eqs. (18, 19)) and the measured optical properties of the metamaterial is demonstrated by ellipsometry (Fig. 15a) and polarization reflectometry (Fig. 15b) of the samples.

Variable angle spectroscopic ellipsometry with photon energies between 0.6 eV and 6.5 eV on the Al/Al$_2$O$_3$ metamaterial have been performed using a Woollam Variable Angle Spectroscopic Ellipsometer (W-VASE). For a uniaxial material with optic axis perpendicular to the sample surface and in plane of incidence, ellipsometry provides the pseudo-dielectric function which, in general, depends both on $\varepsilon_1$ and $\varepsilon_2$. However, as demonstrated by Jellison and Baba [32], the pseudo-dielectric function in this measurement geometry is dominated by the in-plane dielectric function $\varepsilon_1$ and is independent of the angle of incidence. The measured results for the real and imaginary parts of the pseudo-dielectric function in Fig. 15a show good agreement with the model for the in-plane dielectric function (Eq. (18)). The calculated data points are based on the real and imaginary parts of $\varepsilon_{Al}$ tabulated in [30]. The measured sign of the real part of the pseudo-dielectric function is negative, which suggests metallic in-plane transport. The sign of the real part of $\varepsilon_2$ (and therefore, the hyperbolic character of our samples) was determined by polarization reflectometry, since ellipsometry data are less sensitive to $\varepsilon_2$ [32]. Polarization reflectometry also confirmed the negative sign of the real part of $\varepsilon_1$ consistent with ellipsometry data. The metamaterial parameters were extracted from the polarization reflectometry data as described in detail in [33]. Reflectivity for s-polarization is given in terms of the incident angle $\theta$ by

$$R_s = \left| \frac{\sin(\theta - \theta_{ts})}{\sin(\theta + \theta_{ts})} \right|^2 , \qquad (20)$$

where

$$\theta_{ts} = \arcsin\left(\frac{\sin\theta}{\sqrt{\varepsilon_1}}\right).$$

(21)

Reflectivity for p-polarization is given as

$$R_p = \left|\frac{\varepsilon_1 \tan\theta_{tp} - \tan\theta}{\varepsilon_1 \tan\theta_{tp} + \tan\theta}\right|^2,$$

(22)

where

$$\theta_{tp} = \arctan\sqrt{\frac{\varepsilon_2 \sin^2\theta}{\varepsilon_1\varepsilon_2 - \varepsilon_1 \sin^2\theta}}.$$

(23)

We measured p- and s- polarized absolute reflectance on the metamaterial sample using the reflectance mode of the ellipsometer. The reflectance was measured at two photon energies, 2.07 eV (600 nm) and 2.88 eV (430 nm), as shown in Fig. 15b and was normalized to the measured reflectance of a 150 nm gold film. The absolute reflectance of the gold film was obtained from ellipsometry measurements. The estimated uncertainty in the absolute reflectance of the $Al/Al_2O_3$ metamaterial is one percent. In order to obtain the dielectric permittivity $\varepsilon_1$ and $\varepsilon_2$ values, we fit the s- polarized reflectance first, and get the in-plane dielectric function $\varepsilon_1$. We then use the in-plane dielectric function to fit the p- polarized reflectance to obtain the out-of-plane dielectric function, $\varepsilon_2$. The data analysis was done using W-VASE software. At 2.07 eV (600 nm), $\varepsilon_1 = -7.17+1.86i$ and $\varepsilon_2 = 1.56+0.21i$, and at 2.88 eV (430 nm), $\varepsilon_1 = -2.15+0.50i$ and $\varepsilon_2 = 1.30+0.09i$. It is clear that the real part of the out-of-plane dielectric function is positive while the real part of the in-plane dielectric function is negative, which confirms the dielectric nature along z-axis and metallic nature in the xy-plane i.e. a hyperbolic metamaterial.

The $T_c$ and critical magnetic field, $H_c$, of various samples (Figs. 16,17) were determined via four-point resistivity measurements as a function of temperature and magnetic field, H, using a Physical Property Measurement System (PPMS) by Quantum Design. Even though the lowest achievable temperature with our PPMS system was 1.75 K, which is higher than the critical temperature $T_{cAl} = 1.2$ K of bulk aluminum, we were able to observe a pronounced effect of the number of layers on $T_c$ of the hyperbolic metamaterial samples. Fig. 16a shows measured resistivity as a function of temperature for the 1-layer, 3-layer and 8-layer samples each having the same 8.5 nm layer thickness. While the superconducting transition in the 1-layer sample was below 1.75 K, and could not be observed, the 3-layer and 8-layer metamaterial samples exhibited progressively higher critical temperature, which strongly indicates the role of hyperbolic geometry in $T_c$ enhancement. A similar set of measurements performed for several samples having 13 nm layer thickness is shown in Fig. 16b.

The measurements of $H_c$ in parallel and perpendicular fields are shown in Fig. 17. Fig. 17a shows measured resistivity as a function of temperature for a 16-layer 13.2 nm layer thickness hyperbolic metamaterial sample. The critical temperature of this sample appears to be $T_c = 2.3$ K, which is about two times higher than the $T_c$ of bulk aluminum (another transition at $T_c = 2.0$ K probably arise from one or two decoupled layers or edge shadowing effects where the thickness of the films is not uniform). The inset in Fig. 17a illustrates the measurements of $H_c^{parallel}$ for this sample. The critical field appears to be quite large (~3T), which is similar to the values of $H_c^{parallel}$ observed previously in granular aluminium films [34]. However, it is remarkable that such high critical parameters are observed for the films, which are much thicker than granular Al films.

Measurements of the perpendicular critical field $H_c{}^{\text{perp}}$ for the same metamaterial sample, which are shown in Fig. 17b allowed us to evaluate the Pippard coherence length

$$\xi = \sqrt{\frac{\phi_0}{2\pi H_{c2}}}$$

(24)

Assuming $H_{c2}{}^{\text{perp}} = 100$ G (based on the inset in Fig.17b) the corresponding coherence length appears to be $\xi = 181$ nm, which is much larger than the layer periodicity. Other measured samples also exhibit the coherence length around 200 nm. Therefore, our use of effective medium approach is validated and our multilayer samples should obey the metamaterial theory.

We have also studied changes in $T_c$ as a function of Al layer thickness in a set of several 8-layer Al/Al$_2$O$_3$ metamaterial samples, as shown in Fig. 18a. The quantitative behaviour of $T_c$ as a function of $n$ may be predicted based on the hyperbolic enhancement of the electron-electron interaction (Eq. (2)) and the density of electronic states, $\nu$, on the Fermi surface which experience this hyperbolic enhancement. Using Eqs. (18, 19), the effective Coulomb potential from Eq. (2) may be re-written as

$$V(\vec{q}, \omega) = \frac{4\pi e^2}{q^2 \left( \dfrac{q_z^2}{q^2} \dfrac{\varepsilon_d \varepsilon_m}{\left( (1-n)\varepsilon_m + n\varepsilon_d \right)} + \dfrac{q_x^2 + q_y^2}{q^2} \left( n\varepsilon_m + (1-n)\varepsilon_d \right) \right)} \ .$$

(25)

Let us assume once again (see ref.[15] and the theoretical discussion in the previous section) that the dielectric response function of the metal used to fabricate the hyperbolic metamaterial shown in Fig. 11 may be described by Eq.(5), and the critical temperature of a metamaterial superconductor may be found using Eq.(6) via the density of states in Eq.(7).

Let us consider the region of four-momentum $(q, \omega)$ space, where $\omega > \Omega_1(q)$. While $\varepsilon_m = 0$ at $\omega = \Omega_1(q)$, $\varepsilon_m(q, \omega)$ is large in a good metal and negative just above $\Omega_1(q)$. Based on Eq. (25), the differential of the product $\mathcal{W}$ may be written as

$$d(\mathcal{W}) = \frac{4\pi e^2 n \sin\theta \, d\theta}{q^2 \left( \frac{\varepsilon_d \varepsilon_m}{\left((1-n)\varepsilon_m + n\varepsilon_d\right)} \cos^2\theta + \left(n\varepsilon_m + (1-n)\varepsilon_d\right)\sin^2\theta \right)} =$$

$$= -\frac{4\pi e^2 n \, dx}{q^2 \left( \left(n\varepsilon_m + (1-n)\varepsilon_d\right) - \frac{n(1-n)\left(\varepsilon_m - \varepsilon_d\right)^2}{\left(n\varepsilon_d + (1-n)\varepsilon_m\right)} x^2 \right)} \; , \tag{26}$$

where $x = \cos\theta$, and $\theta$ varies from 0 to $\pi$. The latter expression has two poles at

$$\varepsilon_m = \left( \left(1 - \frac{1}{2n(1-n)(1-x^2)}\right) \pm \sqrt{\left(1 - \frac{1}{2n(1-n)(1-x^2)}\right)^2 - 1} \right)\varepsilon_d \tag{27}$$

As the volume fraction, $n$, of metal is varied, one of these poles remains close to $\varepsilon_m = 0$, while the other is observed at larger negative values of $\varepsilon_m$:

$$\varepsilon_m^+ \approx -n(1-n)(1-x^2)\varepsilon_d \tag{28}$$

$$\varepsilon_m^- \approx -\frac{\varepsilon_d}{n(1-n)(1-x^2)} \tag{29}$$

This situation is similar to calculations of $T_c$ for ENZ metamaterials. Since the absolute value of $\varepsilon_m$ is limited (see Eq. (5)), the second pole disappears near $n = 0$ and near $n = 1$. Due to the complicated angular dependences in Eq. (27), it is convenient to reverse the order of integration in Eq. (7), and perform the integration over $d\omega$ first, followed by angular averaging. Following the commonly accepted approach, while integrating over $d\omega$ we take into account only the contributions from the poles given by Eq. (27), and assume the value of $\text{Im}\,\varepsilon_m = \varepsilon_m''$ to be approximately the same at both poles. The respective contributions of the poles to $d(\nu W)/dx$ may be written as

$$\frac{d(\nu W)}{dx} \approx \frac{2\pi e^2 \left( (1-n)\left( \left( 1 - \frac{1}{2n(1-n)(1-x^2)} \right) \pm \sqrt{ \left( 1 - \frac{1}{2n(1-n)(1-x^2)} \right)^2 - 1 } \right) + n \right)}{q^2 \varepsilon''_m (1-n)(1-x^2) \sqrt{ \left( 1 - \frac{1}{2n(1-n)(1-x^2)} \right)^2 - 1 }} \qquad (30)$$

Near $n=0$ and $n=1$ these expression may be approximated as

$$\frac{d(\nu W)^+}{dx} \approx \frac{4\pi e^2 n^2}{q^2 \varepsilon''_m} \qquad (31)$$

and

$$\frac{d(\nu W)^-}{dx} \approx \frac{4\pi e^2}{q^2 \varepsilon''_m (1-x^2)} \ , \qquad (32)$$

respectively. Note that at the $\omega = \Omega_1(q)$ zero of the dielectric response function of the bulk metal the effective Coulomb potential inside the metal may be approximated as

$$V_m(\vec{q}, \omega) = \frac{4\pi e^2}{q^2 \varepsilon_m(q, \omega)} \approx -\frac{4\pi e^2}{q^2 \varepsilon''_m} \ , \qquad (33)$$

so that the coupling constant $\lambda_{\text{eff}}$ of the hyperbolic metamaterial obtained by angular integration of the sum of Eqs. (31) and (32) may be expressed via the coupling constant $\lambda_m$ of the bulk metal:

$$\lambda_{eff} \approx \lambda_m \left( n^2 + \alpha \ln \left| \frac{1 + x_0}{1 - x_0} \right| \right), \qquad (34)$$

where $\alpha$ is a constant of the order of 1 and $x_0$ is defined by the maximum negative value of $\varepsilon_m$, which determines if the second pole (Eq. (29)) exists at a given $n$. Based on Eq.(29),

$$x_0^2 \approx 1 + \frac{\varepsilon_d}{n(1-n)\varepsilon_{m,\max}} \qquad (35)$$

If the second pole does not exist then $x_0 = 0$ may be assumed. Based on Eq. (6), the theoretically predicted value of $T_c$ for the hyperbolic metamaterial is calculated as

$$T_c = T_{Cbulk}\exp\left(\frac{1}{\lambda_m}-\frac{1}{\lambda_{eff}}\right)=T_{Cbulk}\exp\left(\frac{1}{\lambda_m}\left(1-\frac{1}{n^2+\alpha\ln\left|\frac{1+x_0}{1-x_0}\right|}\right)\right) \qquad (36)$$

assuming the known values $T_{cbulk} = 1.2$ K and $\lambda_m = 0.17$ for bulk aluminum [15]. The predicted behaviour of $T_c$ as a function of $n$ is plotted in Fig. 18b. This figure demonstrates that the experimentally measured behaviour of $T_c$ as a function of $n$ (which is defined by the Al layer thickness) correlates well with the theoretical fit, which was obtained using Eq. (36) based on the hyperbolic mechanism of $T_c$ enhancement.

The observed combination of transport and critical properties of the Al/Al$_2$O$_3$ hyperbolic metamaterials is very far removed from the parameter space typical of the granular aluminum films [28, 34]. Together with the number of layer and layer thickness dependences of $T_c$ and $H_c$ shown in Figs. 16-18, these observations strongly support the hyperbolic metamaterial mechanism of superconductivity enhancement. The developed technology enables efficient nanofabrication of thick film aluminum-based hyperbolic metamaterial superconductors with a $T_c$ that is two times that of pure aluminum and with excellent transport and magnetic properties. While the observed $T_c$ increase is slightly smaller than the one observed in ENZ metamaterials [4], the hyperbolic metamaterial geometry exhibits superior transport and magnetic properties compared to the ENZ core-shell metamaterial superconductors. In addition, our theoretical model is applicable to previous experiments performed in NbN/AlN [35] and Al/Si [36] multilayer geometries. We should also note that unlike recent pioneering work on quantum metamaterials [37], which are based on superconducting split-ring resonators and quantum circuits, our work aims at engineering of metamaterials with enhanced superconducting properties.

## 5. What next? How to further increase critical temperature in a metamaterial superconductor.

While theoretical plots in Fig. 1b unambiguously identify the physical mechanism of $T_c$ enhancement in the Al-Al$_2$O$_3$ core-shell metamaterial superconductor, it must be understood that a complete theory should take into account simultaneous contributions of both poles of Eq. (10) to $\lambda_{\text{eff}}$ of the metamaterial. It is also clear that according to Eq. (5) the metamaterial pole occurs at a slightly different frequency compared to the $\varepsilon_m \approx 0$ pole, and therefore it must have slightly different value of $\varepsilon_m'' = \text{Im}\,\varepsilon_m$ compared to the value of $\varepsilon_m''$ for pure aluminium at $\text{Re}\,\varepsilon_m \approx 0$. In order to obtain more precise values, let us consider the behaviour of $\varepsilon_m$ in more detail. Since

$$\left(1 - \frac{\Omega_1^2(q)}{\omega^2 + i\omega\Gamma_1}\right) = \frac{(\omega + \Omega_1)(\omega - \Omega_1) + i\omega\Gamma_1}{\omega^2 + i\omega\Gamma_1}, \tag{37}$$

let us assume that near $\omega = \Omega_1(q)$ Eq. (5) may be approximated as

$$\varepsilon_m \approx -E_m\left(\omega - \Omega_1\right) + i\varepsilon_m^{''}(\omega) \quad , \tag{38}$$

where $E_m$ is a positive constant. Based on Eq. (11), the corresponding frequency of the "metamaterial" pole is

$$\omega = \Omega_1 + \frac{(3-2n)\varepsilon_d}{2nE_m}, \tag{39}$$

which is slightly higher than $\Omega_1$, and therefore a slightly larger value of $\text{Im}(\varepsilon_m) = \varepsilon_m^{''}\left(\Omega_1 + (3-2n)\varepsilon_d / 2nE_m\right)$ may be expected at the metamaterial pole at this higher frequency. As a result, by taking into account simultaneous contributions from both poles (by adding the contributions to $\lambda_{eff}$ from the "metal" and the "metamaterial" poles given by Eqs.(13) and (15), respectively), the final expression for $\lambda_{\text{eff}}$ is

$$\lambda_{eff} \approx \frac{n^2}{(3-2n)}\lambda_m + \frac{9(1-n)}{2(3-2n)}\frac{\varepsilon_m^{''}}{\varepsilon_{mm}^{''}}\lambda_m = \lambda_m\left(\frac{n^2}{(3-2n)} + \frac{9(1-n)\alpha}{2(3-2n)}\right), \tag{40}$$

where $\alpha = \varepsilon_m^{''}/\varepsilon_{mm}^{''} < 1$ is determined by the dispersion of $\varepsilon_m^{''}$. Substitution of Eq. (40) into Eq. (16) produces the following final expression for the critical temperature of the metamaterial:

$$T_c = T_{Cbulk}\exp\left(\frac{1}{\lambda_m}\left(1 - \frac{1}{\left(\frac{n^2}{(3-2n)} + \frac{9(1-n)\alpha}{2(3-2n)}\right)}\right)\right) \text{ at } n > n_{\text{cr}}, \tag{41}$$

which now has a single fitting parameter $\alpha$. We will consider it as a free parameter of the model, since we are not aware of any experimental measurements of dispersion of $\text{Im}\varepsilon_m$ for aluminum. Note that Eq. (41) depends on $\varepsilon_d$ only via $n_{\text{cr}}$ and $\alpha$. The calculated behavior of $T_c$ as a function of $n$ is presented in Fig.19a. The agreement appears to be very good. The best fit to experimental metamaterial data is obtained at $\alpha = 0.89$. We must emphasize that this model is quite insensitive to the particular choice of functional form of $\varepsilon_m(q, \omega)$ in the broad $(q, \omega)$ range. In our derivations we are only using the fact that $\varepsilon_m$ changes sign near $\omega = \Omega_1(q)$ by passing through zero, which is described by Eq.(38).

The apparent success of such a simple theoretical description in the case of Al-$Al_2O_3$ core-shell and hyperbolic metamaterial superconductors prompted us to apply the same theory to the tin-based ENZ metamaterials studied in [3]. Assuming the known values $Tc_{bulk}$=3.7K and $\theta_D = 200$ K of bulk tin [17,18], Eq. (6) results in $\lambda_m = 0.25$, which also corresponds to the weak coupling limit. The calculated behavior of $T_c$ as a function of $n$ is presented in Fig. 19b, which also shows experimental data points measured in [3]. The agreement also appears to be very good, given the spatial inhomogeneity of the fabricated metamaterials [3] (note also the difference in temperature scales in Figs. 19a and 19b). The best fit to experimental metamaterial data is obtained at $\alpha = 0.83$. It appears that a smaller increase in metamaterial $T_c$ in this case is due to larger value of permittivity $\varepsilon_d$ of the dielectric component of the metamaterial, which according to Eq. (11) limits the range $n_{cr} < n < 0.6$ where the metamaterial $T_c$ enhancement may occur. Larger value of $\varepsilon_d$ also leads to smaller value of $\alpha$ (see Eq. (39)), which also reduces the metamaterial pole contribution to $\lambda_{eff.}$

We should also note that the appearance of an additional "metamaterial" pole in the Maxwell-Garnett expression (Eq. (9)) for a metal-dielectric mixture does not rely on any particular spatial scale, as long as it is smaller than the coherence length. In fact, the Maxwell-Garnett approximation may be considered as a particular case of the Clausius–Mossotti relation [21], which traces the dielectric constant of a mixture to the dielectric constants of its constituents, and it is only sensitive to the volume fractions of the constituents. Therefore, the same concept is supposed to be applicable at much smaller spatial scales compared to the two metamaterial cases studied in Refs. 3 and 4. We anticipate that similar $T_c$ enhancements may be observed in metamaterials based on other higher temperature superconductors, which have smaller coherence length, such as niobium and $MgB_2$.

First, let us consider the case of niobium, which has $Tc_{bulk} = 9.2$ K and $\theta_D = 275$K in the bulk form [17,18]. Enhancing the superconducting properties of niobium is a very important task, since niobium alloys, such as $Nb_3Sn$ and NbTi are widely used in superconducting cables and magnets. Based on Eq. (6) niobium has $\lambda_m = 0.29$, which also corresponds to the weak coupling limit. The coherence length of niobium is $\xi = 38$ nm [17,18], which would complicate nanofabrication requirements for a niobium-based metamaterial. However, given the current state of nanolithography, which operates on a 14 nm node [38], these requirements still look quite realistic. Alternatively, smaller than 38 nm diameter nanoparticles would need to be used in nanoparticle-based ENZ metamaterial geometries.

The calculated behavior of $T_c$ of a niobium-based ENZ metamaterial as a function of $n$ is presented in Fig. 20a for different values of the $\alpha$ ratio in Eq. (41). The vertical dashed line in the plot corresponds to the same value of $n_{cr}$ as in the $Al$-$Al_2O_3$ core-shell metamaterial superconductor. It is interesting to note that the general range of predicted $T_c$ enhancement matches well with the observed enhancement of $T_c$ in various niobium alloys [17]. The corresponding experimental data point for $Nb_3Sn$ is shown in the same plot for comparison (it is assumed that $n = 0.75$ for $Nb_3Sn$). As we have mentioned above, our model is based on the Maxwell-Garnett approximation, which traces the dielectric constant of a mixture to the dielectric constants of its constituents.

The spatial scale of mixing is not important within the scope of this model. Therefore, our model may be applicable to some alloys. We should note that the additional "metamaterial" pole described by Eq. (11) may also be present in the case of a mixture of a normal metal with the superconductor. However, unlike the previously considered case of a dielectric mixed with a superconductor, the additional "metamaterial" pole is observed at $\omega < \Omega_1$ where the dielectric constant of the superconductor $\varepsilon_s$ is positive (see Eq. (38)), while the dielectric constant of the normal metal $\varepsilon_m$ is negative (it is assumed that the critical temperature of the normal metal is lower than the critical temperature of the superconductor). Indeed, under these conditions an additional pole in the inverse effective dielectric response function of the metamaterial does appear in (Eq. (10)), and similar to Eq. (11), it is defined as

$$\varepsilon_s \approx -\frac{3-2n}{2n}\varepsilon_m \qquad (42)$$

This pole will result in the enhanced $T_c$ of the metamaterial (or alloy), which may be calculated using the same Eq. (41). However, since this additional pole occurs at $\omega < \Omega_1$, it is expected that Im$\varepsilon_s$ will be smaller at this frequency, resulting in factor $\alpha$ being slightly larger than 1. This conclusion agrees well with the position of Nb$_3$Sn data point on Fig.20a.

Next, let us consider the case of MgB$_2$, which has Tc$_{bulk}$ = 39 K and $\theta_D$ = 920 K in the bulk form [39]. The coherence length of MgB$_2$ remains relatively large: the $\pi$ and $\sigma$ bands of electrons have been found to have two different coherence lengths, 51 nm and 13 nm [40]. Both values are large enough to allow metamaterial fabrication, at least in principle. Based on Eq. (6) MgB$_2$ has $\lambda_m$ = 0.32, which still remains within the scope of the weak coupling limit. The calculated behavior of $T_c$ of a MgB$_2$-based ENZ metamaterial as a function of $n$ is presented in Fig.4b for different values of the $\alpha$ ratio in Eq. (41). Similar to Fig. 20a, the vertical dashed line in the plot corresponds to the same value of $n_{cr}$ as in the Al-Al$_2$O$_3$ core-shell metamaterial superconductor. It appears that the critical temperature of MgB$_2$-based ENZ metamaterial would probably fall in the liquid nitrogen temperature range. A good choice of the dielectric component of such a metamaterial could be diamond: 5 nm diameter diamond nanoparticles are available commercially – see for example Ref. 3.

Unfortunately, superconductors having progressively higher $T_c$ are not well described by Eq. (6), which is valid in the weak coupling limit only. So it is interesting to evaluate what kind of metamaterial enhancement may be expected in such system as H$_2$S, which superconducts at 203 K at very high pressure [41]. While simple theoretical description developed in our work may not be directly applicable to this case as far as $T_c$ calculations are concerned (because of the very large values of $\theta_D \sim 1200$ K and $\lambda \sim 2$ in this material [42]) adding a suitable dielectric or normal metal to H$_2$S will still result in an additional "metamaterial" pole in its inverse dielectric response function. Eqs.(8- 12) will remain perfectly applicable in this case. Therefore, adding such a pole will enhance superconducting properties of H$_2$S. Based on Eq. (10) we may predict that the enhancement of superconducting properties will be observed at $n \sim 0.6$, as it also happened in the cases of aluminum and tin-based metamaterials considered in this paper. As far as the choice of a dielectric is concerned, based on Eqs. (8-12) it appears that readily available nanoparticles of diamond (off the shelf 5 nm diameter diamond nanoparticles were used in [3]) could be quite suitable, since diamond has rather low dielectric constant $\varepsilon_{dia} \sim 5.6$, which stays almost constant from the visible to RF frequency range. Low permittivity of the dielectric component ensures smaller value of $n_{cr}$ defined by Eq.(11), and hence larger metamaterial enhancement. It is not inconceivable that 5 nm diameter diamond nanoparticles could be added to $H_2S$ in the experimental chamber used in [41].

Unfortunately, in the strong coupling limit the expected $T_c$ enhancement may not be as drastic as in the weak coupling limit described by Eq. (6). In the $\lambda >> 1$ limit of the Eliashberg theory the critical temperature of the superconductor may be obtained as [43]

$$T_c = 0.183\theta_D \sqrt{\lambda} \qquad (43)$$

Therefore, enhancement of $\lambda$ by the largest possible factor of 1.5 calculated based on the Maxwell-Garnett theory for the effective dielectric response function of the metamaterial (see Fig.1a) will lead to $T_c$ enhancement by factor of $(1.5)^{1/2}$, which means that a critical temperature range $\sim 250$ K (or $\sim$ -20$^o$ C) may potentially be achieved in a $H_2S$ based metamaterial superconductor. Such a development would still be of sufficient interest, since this would constitute an almost room temperature superconductivity.

Considerable increase of Tc of the metamaterial superconductors may also be achieved by optimizing the metamaterial geometry. Historically, the suggestion to use highly polarizable dielectric side chains or layers in order to increase $T_c$ of a 1D or 2D electronic system can be traced back to the pioneering papers by W. Little [44] and V. Ginzburg [45]. Using modern language, the superconducting systems proposed in these works could be called "superconducting metamaterials". While these proposals never led to hypothesized room temperature superconductivity, experimental attempts to realize such superconductors were limited by modest polarizability of natural materials. Recent development of high $\varepsilon$ metamaterials gives these proposals another chance [2]. An example of the high-index metamaterial, which may be used in the modern day versions of either 1D or 2D superconducting geometries proposed by Little and Ginzburg is shown in Fig.21 [2].

In another recent development, fractal metamaterial superconductor geometry has been suggested and analyzed based on the theoretical description of critical temperature increase in epsilon near zero (ENZ) metamaterial superconductors [46]. Considerable enhancement of critical temperature has been predicted in such fractal metamaterials due to appearance of large number of additional poles in the inverse dielectric response function of the fractal. Indeed, relatively large value of the superconducting coherence length $\xi = 1600$ nm of bulk aluminum [17] leaves a lot of room for engineering an aluminum-based fractal metamaterial superconductor according to the procedure shown schematically in Fig.22a.

Let us consider the central image (the 2nd fractal order) from Fig. 22a. We may use the Maxwell-Garnett approximation (Eq. (8)) in order to determine the effective dielectric permittivity of such metamaterial. However, this time we will assume that the "first order metamaterial" described by Eq. (9) plays the role of the "matrix". The volume fraction of the new matrix is $n$, and it is diluted with the same dielectric $\varepsilon_d$ with volume fraction $(1-n)$. Mixing of these components results in the effective medium with a dielectric constant $\varepsilon^{(2)}_{eff}$, which may be obtained as (compare to Eq. (9)):

$$\varepsilon^{(2)}_{eff} = \frac{\varepsilon^{(1)}_{eff}\left((3-2n)\varepsilon_d + 2n\varepsilon^{(1)}_{eff}\right)}{\left(n\varepsilon_d + (3-n)\varepsilon^{(1)}_{eff}\right)} \qquad (44)$$

It is easy to verify that in addition to the pole at $\varepsilon_m \approx 0$ and the first order pole described by Eq. (11), the inverse dielectric response function of the "second order" fractal metamaterial has additional poles at

$$\varepsilon_m = -\varepsilon_d \, \frac{(3-2n)(3+n)}{8n^2} \left( 1 \pm \sqrt{1 - \frac{16n^3}{(3-2n)(3+n)^2}} \right) \qquad (45)$$

These additional poles may be expressed in the recurrent form as

$$\varepsilon_m^{(2)} = \varepsilon_m^{(1)} \, \frac{(3+n)}{4n} \left( 1 \pm \sqrt{1 - \frac{16n^3}{(3-2n)(3+n)^2}} \right) \qquad (46)$$

The latter equation, which expresses $\varepsilon_m^{(k+1)}$ as a function of $\varepsilon_m^{(k)}$ may be used to determine all the poles of the fractal metamaterial structure using an iterative procedure. The first few poles determined in such a way are plotted in Fig. 22b as a function of $n$ for the case of $\varepsilon_d = 3$. This plot makes it clear that most of the poles in each iteration (which correspond to the minus sign in Eq. (46)) stay near $\varepsilon_m \approx 0$, while one pole in each iteration (corresponding to the plus sign) may be written as

$$\varepsilon_m^{(k+1)} \approx \varepsilon_m^{(k)} \, \frac{(3+n)}{2n} \qquad (47)$$

The vertical dashed line in Fig. 22b corresponds to the critical value of the metal volume fraction $n_{cr}$ observed in the case of the Al-Al$_2$O$_3$ core-shell superconductor [3]. As indicated in Fig. 22b, this value may be used as a guideline to evaluate how many additional poles may be observed in a real fractal metamaterial at a given volume fraction $n$. The additional poles of the inverse dielectric response function of the fractal metamaterial defined by Eq. (47) may be described as the plasmon-phonon modes of the metamaterial, which participate in the Cooper pairing of electrons in a fashion, which is quite similar to the conventional BCS mechanism.

Let us evaluate what kind of critical temperature increase may be expected in a fractal metamaterial superconductor because of the appearance of all these additional poles in $\varepsilon^{-1}_{eff}(q, \omega)$. We already have demonstrated that the additional poles of the inverse dielectric response function may be found using an iterative procedure. A similar iterative expression may be derived for $\mathrm{Im}\,\varepsilon^{-1}_{\mathrm{eff}}(q, \omega)$ near these poles. Based on Eq. (10), we may write the following expression for the imaginary part of the inverse dielectric response function of the first order material near the metamaterial pole:

$$\mathrm{Im}\left(\left(\varepsilon^{(1)}_{eff}\right)^{-1}\right) \approx \frac{9(1-n)}{2n(3-2n)}\,\alpha\,\mathrm{Im}\left(\left(\varepsilon^{(0)}_{eff}\right)^{-1}\right), \qquad (48)$$

where $\mathrm{Im}\left(\left(\varepsilon^{(0)}_{eff}\right)^{-1}\right) = 1/\varepsilon^{"}_m$, and $\alpha \sim 1$ near $n \sim 1$. Since the magnitude of $\varepsilon_m$ is finite (see Fig. 22b), in a real fractal metamaterial geometry a large number of poles may be expected only at values of $n$ close to 1. Therefore, from now on we will assume that $\alpha = 1$. An expression similar to Eq. (48) may be applied iteratively to the next fractal orders. As far as the density of states $\nu$ in Eq. (7) is concerned, as illustrated by Fig. 22a, with each iteration the density of states of the metamaterial is multiplied by a factor of $n$. Thus, we may now sum up contributions of all the fractal poles to $\lambda_{\mathrm{eff}}$. It may be obtained as a sum of a geometrical progression:

$$\lambda_{eff} \approx \left( n^{K-1}\left(\frac{9(1-n)}{2(3-2n)}\right) + n^{K-2}\left(\frac{9(1-n)}{2(3-2n)}\right)^2 + \ldots + \left(\frac{9(1-n)}{2(3-2n)}\right)^K \right)\lambda_m =$$

$$= n^{K-1}\left(\frac{9(1-n)}{2(3-2n)}\right) \frac{\left(1 - \left(\frac{9(1-n)}{2n(3-2n)}\right)^K\right)}{\left(1 - \left(\frac{9(1-n)}{2n(3-2n)}\right)\right)} \lambda_m, \qquad (49)$$

assuming that the metamaterial structure contains $K$ fractal orders. The resulting $\lambda_{\mathrm{eff}}$ is plotted in Fig. 23 for the first four fractal orders. The corresponding $n_{cr}^{(k)}$ for each fractal order (determined using Fig. 22b) are marked by the vertical dashed lines in Fig. 23. This figure clearly demonstrates that fractal structure promotes superconductivity by increasing $\lambda_{eff}$. For the case of Al-based metamaterial the enhancement of Tc is expected for at least the first three fractal orders. Compared to a simple metal-dielectric ENZ metamaterial (described as a "first order fractal" in Fig.23), the higher order fractal metamaterial structures exhibit much stronger enhancement of the coupling constant $\lambda_{eff}$, which should lead to much larger $T_c$ increase. Based on Fig. 23, the enhancement of $T_c$ ($\lambda_{eff}/\lambda_m > 1$) in a fractal metamaterial occurs starting at $n \sim 0.8$, which is close enough to $n \sim 1$ to validate our assumptions. The corresponding values of $T_c$ calculated using Eq.(16) based on the superconducting parameters of bulk aluminium for the first two fractal orders are shown in Fig. 24. The calculated curves are terminated at the experimentally defined values of $n_{cr}^{(k)}$ for the respective fractal order, which are taken from Fig. 22b. Eight-fold enhancement of $T_c$ compared to the bulk aluminium is projected for the second order fractal structure. The fractal enhancement of Tc may become even more pronounced for materials having larger values of $-\varepsilon_m$ compared to bulk aluminum.

Our results agree with the recent observations [47, 48] that fractal defect structure promotes superconductivity. In particular, it was observed that the microstructures of the transition-metal oxides, including high-$T_c$ copper oxide superconductors, are complex. For example, the oxygen interstitials or vacancies (which strongly influence the dielectric properties of the bulk high-$T_c$ superconductors [49]) exhibit fractal order [47]. These oxygen interstitials are located in the spacer layers separating the superconducting $CuO_2$ planes. They undergo fractal ordering phenomena that induce enhancements in the transition temperatures with no changes in the overall hole concentrations. Such ordering of oxygen interstitials in the $La_2O_{2+y}$ spacer layers of $La_2CuO_{4+y}$ high-$T_c$ superconductors is characterized by a fractal distribution up to a maximum limiting size of 400 mm. It was observed [47], quite intriguingly, that these fractal distributions of dopants seem to enhance superconductivity at high temperature, which is difficult to explain since the superconducting coherence length in these compounds is very small $\xi \sim 1 - 2$ nm [47, 48]. Appearance of the plasmon-phonon modes of the fractal structure described above may resolve this issue, since such fractal phonon modes reside on a scale, which is much larger than several nanometers. We should also mention that such a mechanism may also explain observations of increased critical temperatures in fractal Pb thin films [50].

## 6. Conclusion.

Unlike the traditional searches of higher Tc superconductors, which rely on choosing natural materials exhibiting larger electron-electron interactions, the metamaterial superconductor paradigm offers considerably extended range of search scenarios. The newly developed electromagnetic metamaterials offer the field of superconductivity research many novel tools, which enable intelligent engineering of electron pairing interaction. Considerable enhancement of the electron-electron interaction may be expected in such metamaterial scenarios as in epsilon near zero (ENZ) and hyperbolic metamaterials, as well as in novel fractal metamaterial geometries. In all these cases the dielectric function may become small and negative in substantial portions of the relevant four-momentum space, leading to enhancement of the electron pairing interaction. This approach has been verified in experiments with aluminium-based metamaterials. Metamaterial superconductor with Tc = 3.9 K have been fabricated, that is three times that of pure aluminium (Tc = 1.2 K), which opens up new possibilities to considerably improve Tc of other simple superconductors. Taking advantage of the demonstrated success of this approach, the critical temperature of hypothetic niobium,

MgB2 and H2S-based metamaterial superconductors has been evaluated. The MgB2-based metamaterial superconductors are projected to reach the liquid nitrogen temperature range. In the case of an H2S-based metamaterial projected Tc appears to be as high as ~250 K, which almost reaches the room temperature range. We hope that our brief review clearly demonstrates that the future of metamaterial superconductors is bright, and we anticipate many novel exciting scientific and technical applications of these materials.

**Acknowledgement**

This work was supported in part by the DARPA Award No: W911NF-17-1-0348 "Metamaterial Superconductors".

**Figure Captions**

**Figure 1**. (a) Predicted behaviour of $\lambda_{eff}/\lambda_m$ as a function of metal volume fraction $n$ in a metal-dielectric metamaterial. (b) Plots of the hypothetic values of $T_c$ as a function of metal volume fraction $n$, which would originate from either Eq. (13) or Eq. (15) in the absence of each other. The experimentally measured data points from [4] are shown for comparison on the same plot. The vertical dashed line corresponds to the assumed value of $n_{cr}$. Note that the theoretical curves contain no free parameters.

**Figure 2.** Numerical calculations of the real part of the dielectric constant of the $Sn/BaTiO_3$ mixture as a function of volume fraction of $BaTiO_3$ performed with 10% steps.

**Figure 3.** (a) SEM image of the composite $Sn/BaTiO_3$ metamaterial with 30% volume fraction of $BaTiO_3$ nanoparticles. Individual compressed nanoparticles are clearly visible in the image. (b) Schematic diagram of the metamaterial sample geometry.

**Figure 4**. Temperature dependence of zero field cooled magnetization per unit mass for several samples with varying concentration of (a) $BaTiO_3$ and (b) $SrTiO_3$ measured in magnetic field of 10 G.

**Figure 5.** The temperatures of the onset of the superconducting transition (blue squares) and the temperatures of the midpoint of the transition (red circles) plotted as a function of volume fraction of $BaTiO_3$. Line is a guide to the eye.

**Figure 6**. Schematic geometry of the ENZ metamaterial superconductor based on the core-shell nanoparticle geometry. The nanoparticle diameter is d=18 nm. The inset shows typical dimensions of the fabricated bulk aluminium-based core-shell metamaterial.

**Figure 7**. Comparison of the FTIR reflectivity spectrum of a typical core-shell $Al_2O_3$-Al metamaterial sample with reflectivity spectra of bulk Al and $Al_2O_3$ samples. The step in reflectivity around 11 μm may be used to characterize the volume fraction of $Al_2O_3$ in the core-shell metamaterial. The increased noise near 22 μm is related to the IR source cutoff.

**Figure 8**. The plots of the real part of $\varepsilon$ for pure Al and for the Al-$Al_2O_3$ core-shell metamaterial based on the Kramers-Kronig analysis of the FTIR reflectivity data from Fig.2: (a) Comparison of $\varepsilon$ ' for pure Al and for the Al-$Al_2O_3$ metamaterial clearly indicates that $\varepsilon_{Al\text{-}Al2O3} \ll \varepsilon_{Al}$. (b) Real part of $\varepsilon$ for two different Al-$Al_2O_3$ core-shell metamaterials based on the Kramers-Kronig analysis. While both metamaterials shown in (b) exhibit much smaller $\varepsilon$ compared to the bulk aluminium, the metamaterial prepared using less oxidized aluminium nanoparticles exhibits considerably larger negative $\varepsilon$.

**Figure 9**. The plots of the real part of $\varepsilon$ for pure tin and for the ENZ tin-$BaTiO_3$ nanocomposite metamaterial studied in [3]: (a) Comparison of $\varepsilon$ ' for compressed tin nanoparticles  and for the tin-$BaTiO_3$ nanocomposite metamaterial. (b) Real part of $\varepsilon$ for the tin-$BaTiO_3$ nanocomposite metamaterial.

**Figure 10**. (a) Temperature dependence of zero field cooled magnetization per unit mass for several Al-$Al_2O_3$ core-shell metamaterial samples with increasing degree of oxidation measured in magnetic field of 10 G. The highest onset of superconductivity at ~3.9K is marked by an arrow. This temperature is 3.25 times larger than Tc=1.2K of bulk aluminium.    (b) Corresponding FTIR reflectivity spectra of the core-shell metamaterial samples. Decrease in reflectivity corresponds to decrease of the volume fraction of aluminium.

**Figure 11.** Geometry and basic properties of hyperbolic metamaterial superconductors:

(a) Schematic geometry of a "layered" hyperbolic metamaterial. (b) Electron-electron pairing interaction in a hyperbolic metamaterial is strongly enhanced near the cone in momentum space defined as $q_z^2 \varepsilon_2 + \left( q_x^2 + q_y^2 \right) \varepsilon_1 = 0$.

**Figure 12**. Transmission electron microscope (TEM) image of a 16 layer metamaterial sample. During the imaging experiments samples were coated with gold and platinum to ensure conductivity of the surface during sample preparation. The inset shows that the interfacial layers are amorphous, between polycrystalline grains of Al, that, in cross-section, exhibit Moire cross-hatching.

**Figure 13**. Measurements of DC and AC (optical) conductivity of the aluminum layers of the fabricated hyperbolic metamaterials: (a) Temperature dependences of the sheet resistance of a 16-layer 10nm/layer $Al/Al_2O_3$ hyperbolic metamaterial and a 100 nm thick Al film. As illustrated in the logarithmic plot in the inset, the conductivity of the metamaterial approaches conductivity values of bulk aluminum and is far removed from the parameter space characteristic for granular Al films which is indicated by the gray area. (b) IR reflectivity of bulk aluminium, 3 and 8 layer hyperbolic metamaterial, and the core shell metamaterial samples measured in the long wavelength IR (LWIR) (2.5 - 22.5 μm) range using an FTIR spectrometer. The step in reflectivity around 11 μm is related to the phonon-polariton resonance (PPR) and may be used to characterize the spatial distribution of $Al_2O_3$ in the metamaterial samples.

**Figure 14**. The calculated plots of the real parts of $\varepsilon_{x,y}$ (a) and $\varepsilon_z$ (b) of the multilayer $Al/Al_2O_3$ metamaterial. The metamaterial consists of 13 nm thick Al layers separated by 2 nm of $Al_2O_3$ in the LWIR spectral range. The calculations were performed using Eqs. (18,19) based on the Kramers-Kronig analysis of the FTIR reflectivity of Al and $Al_2O_3$

in ref. [4]. The metamaterial appears to be hyperbolic except for a narrow LWIR spectral band between 11 and 18 μm.

**Figure 15.** Ellipsometry and polarization reflectometry of Al/Al$_2$O$_3$ hyperbolic metamaterials. (a) Comparison of measured pseudo-dielectric function using ellipsometry and theoretically calculated Re$\varepsilon_1$ and Im$\varepsilon_1$. Theoretical data points are based on real and imaginary parts of $\varepsilon_{Al}$ tabulated in [30]. (b) Data points are measured p- and s-polarized reflectivities of the metamaterial sample at 2.07 eV (600 nm) and 2.88 eV (430 nm). Dashed lines are fits using Eq. (20-23). $\varepsilon_1$ and $\varepsilon_2$ obtained from the fits confirm hyperbolic character of the metamaterial.

**Figure 16.** Effect of the number of layers on $T_c$ of the Al/Al$_2$O$_3$ hyperbolic metamaterial samples: (a) Measured resistivity as a function of temperature is shown for the 1-layer, 3-layer and 8-layer samples each having the same 8.5 nm layer thickness. (b) Measured resistivity as a function of temperature for the 1-layer, 3-layer, 8-layer and 16-layer samples each having the same 13 nm layer thickness.

**Figure 17.** Evaluation of the Pippard coherence length of the Al/Al$_2$O$_3$ hyperbolic metamaterial: (a) Measured resistivity as a function of temperature for a 16-layer 13.2 nm layer thickness metamaterial sample. The critical temperature appears to be $T_c$=2.3K. The inset shows resistivity of this sample as a function of parallel magnetic field at T=1.75K. (b) Resistivity of the same sample as a function of perpendicular magnetic field at T=1.75K. Assuming $H_{c2}^{perp}$=100G (based on the measurements shown in the inset) the corresponding coherence length appears to be $\xi$=181 nm, which is much larger than the layer periodicity.

**Figure 18.** Effect of the aluminum volume fraction $n$ on $T_c$ of the Al/Al$_2$O$_3$ hyperbolic metamaterial samples: (a) Resistivity as a function of temperature for the 8-layer

samples having different aluminum layer thicknesses. (b) Experimentally measured behavior of $T_c$ as a function of $n$ (which is defined by the Al layer thickness) correlates well with the theoretical fit (red curve) based on the hyperbolic mechanism of $T_c$ enhancement. Experimental data points shown in black correspond to 8-layer samples, while blue ones correspond to 16-layer samples.

**Figure 19**. (a) Theoretical plot of Tc versus volume fraction n of aluminium in the Al-Al2O3 core-shell metamaterial calculated using Eq. (41). The best fit to experimental data is obtained at $\alpha = 0.89$. (b) Theoretical plot of Tc versus volume fraction n of tin in the tin-BaTiO3 ENZ metamaterial [3] calculated using Eq. (41). The best fit to experimental data is obtained at $\alpha = 0.83$. The experimental data points are taken from Ref. 3.

**Figure 20**. (a) Theoretical plots of $T_c$ versus volume fraction $n$ of niobium in a hypothetic niobium-based ENZ metamaterial calculated using Eq. (41) for different values of the $\alpha$ ratio. The experimental data points for bulk Nb and Nb$_3$Sn are shown in the same plot for comparison. (b) Theoretical plots of $T_c$ versus volume fraction $n$ of MgB$_2$ in a hypothetic MgB$_2$-based ENZ metamaterial calculated using Eq. (41) for different values of the $\alpha$ ratio. In both cases the vertical dashed lines correspond to the same value of $n_{cr}$ as in the Al-Al$_2$O$_3$ core-shell metamaterial superconductor.

**Figure 21**. (a) Unit cell structure of the high-index metamaterial, which may be used in the modern day version of the 1D or 2D superconducting geometries proposed by W. Little [44] and V. Ginzburg [45]. The metamaterial is made of a thin 'I'-shaped metallic patch symmetrically embedded in a dielectric material. The 3D structure of this metamaterial may be described as an array of subwavelength capacitors, in which the gap width defined by $g=L-a$ and the interlayer spacing $d$ (exaggerated for clarity) define the capacitor values. (b) Real and imaginary parts of $\varepsilon_{eff}$ numerically calculated for the metamaterial geometry shown in (a) for the following values of geometrical parameters: $L$=98 nm, $g$=14 nm, $w$=28 nm. The metamaterial exhibits broadband high $\varepsilon$ behaviour, while metamaterial losses remain modest below 50 THz.

**Figure 22**. (a) Schematic geometry of a metal-dielectric fractal metamaterial superconductor. For the first fractal order (the second panel) the volume fraction of the superconductor is $n = 0.56$. (b) The first few poles of the inverse dielectric response function of a fractal metal-dielectric metamaterial plotted as a function of $n$ for the case of $\varepsilon_d = 3$. The fractal order is marked near the curve for each pole. The vertical dashed line corresponds to the critical value of the metal volume fraction $n_{cr}$ observed in the case of the Al-Al$_2$O$_3$ core-shell superconductor. It is used to determine $n_{cr}^{(k)}$ for the next fractal orders, as indicated by the horizontal dashed line.

**Figure 23**. Plot of the magnitude of $\lambda_{eff}$ given by Eq. (49) as a function of $n$ for the first four orders of a fractal metamaterial superconductor. The corresponding $n_{cr}^{(k)}$ are marked by the vertical dashed lines.

**Figure 24**. Enhancement of $T_c$ of the fractal metamaterial superconductor calculated as a function of $n$ based on the superconducting parameters of bulk aluminium. The second order fractal structure demonstrates considerably higher $T_c$ compared to the first order structure.

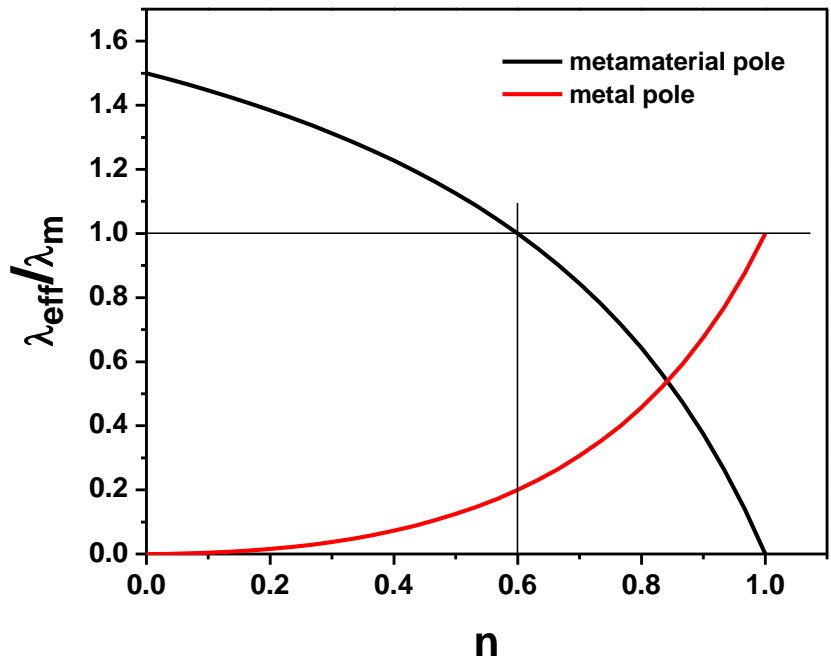

(a)

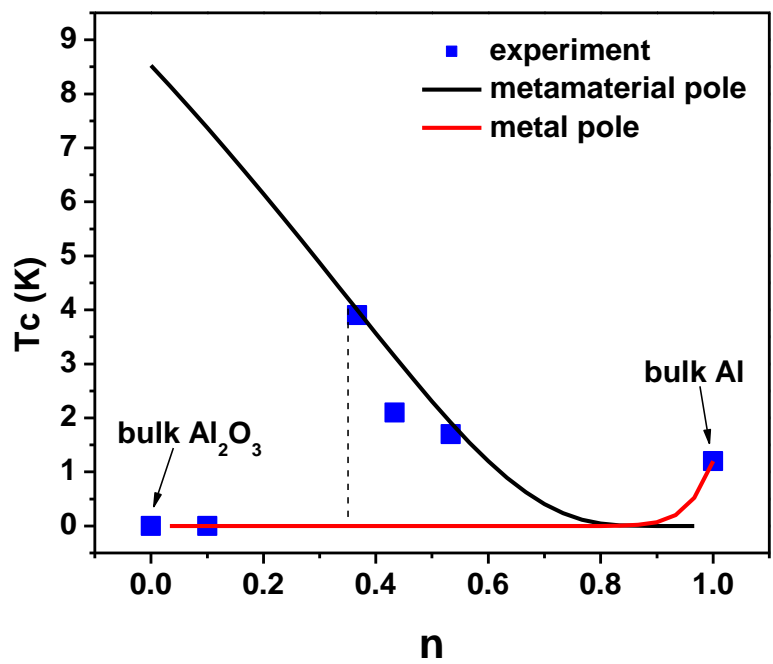

(b)

Fig. 1

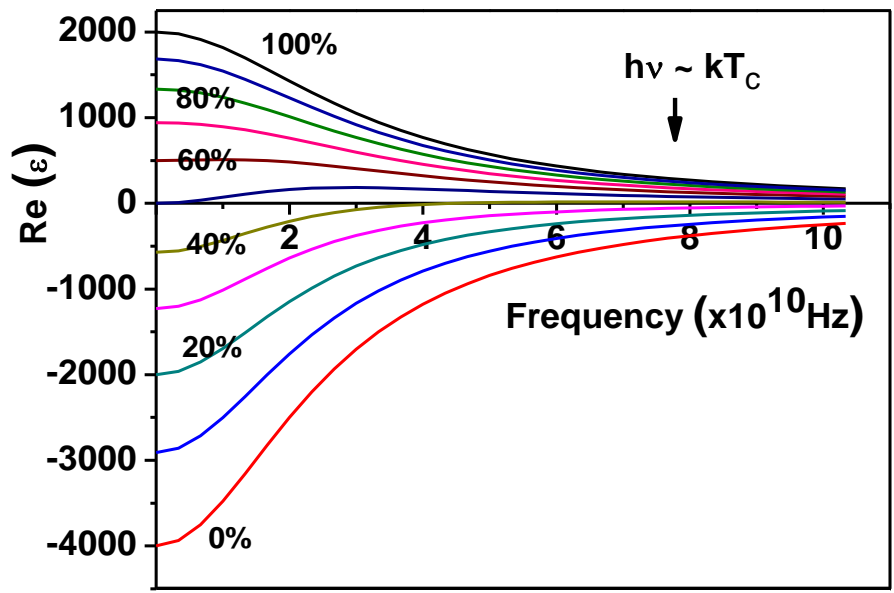

Fig.2

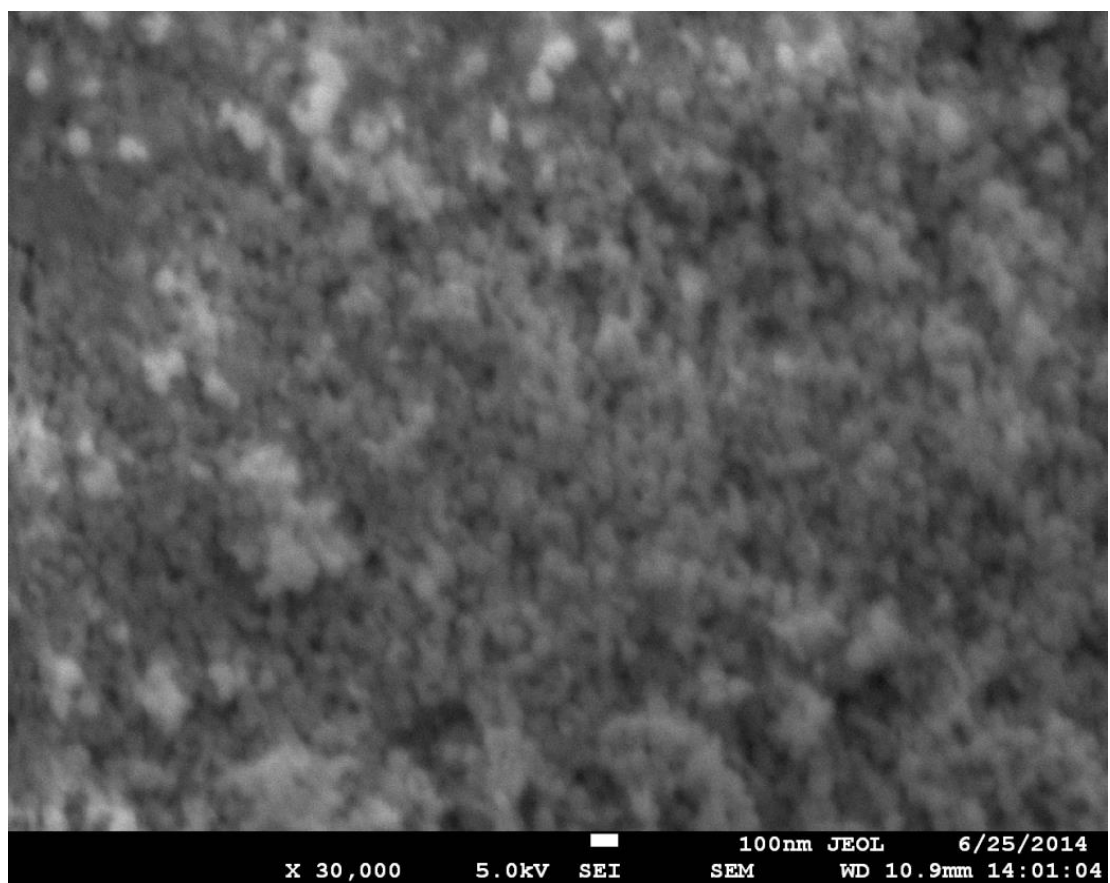

X 30,000  5.0kV  SEI  100nm JEOL  6/25/2014
SEM  WD 10.9mm 14:01:04

(a)

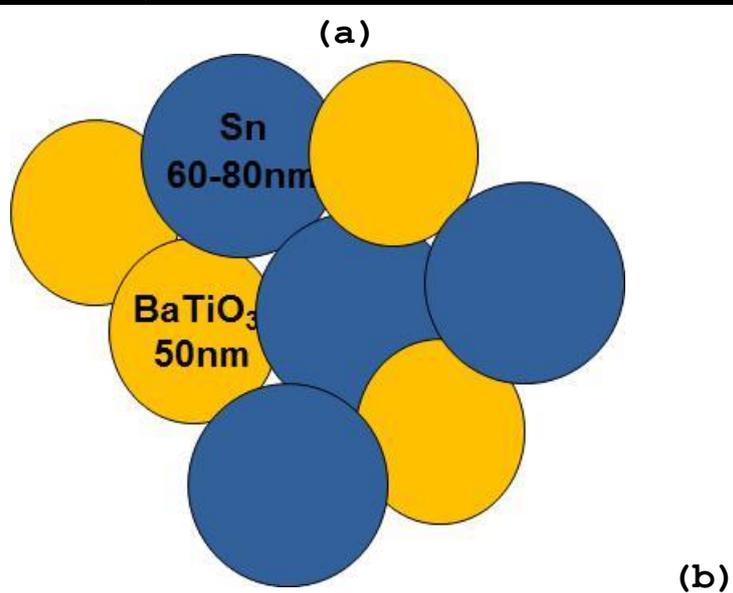

Sn
60-80nm

BaTiO$_3$
50nm

(b)

Fig.3

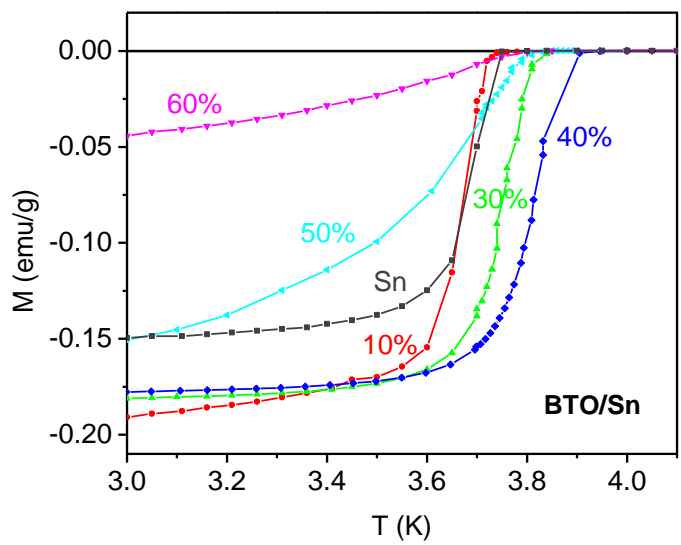

(a)

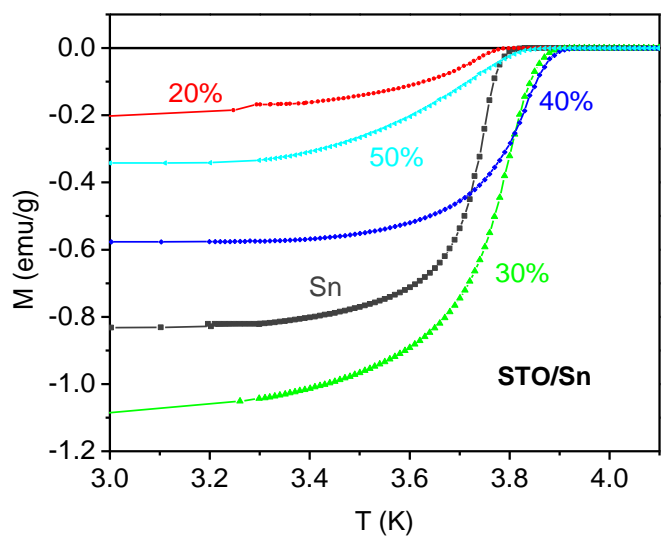

(b)

Fig. 4

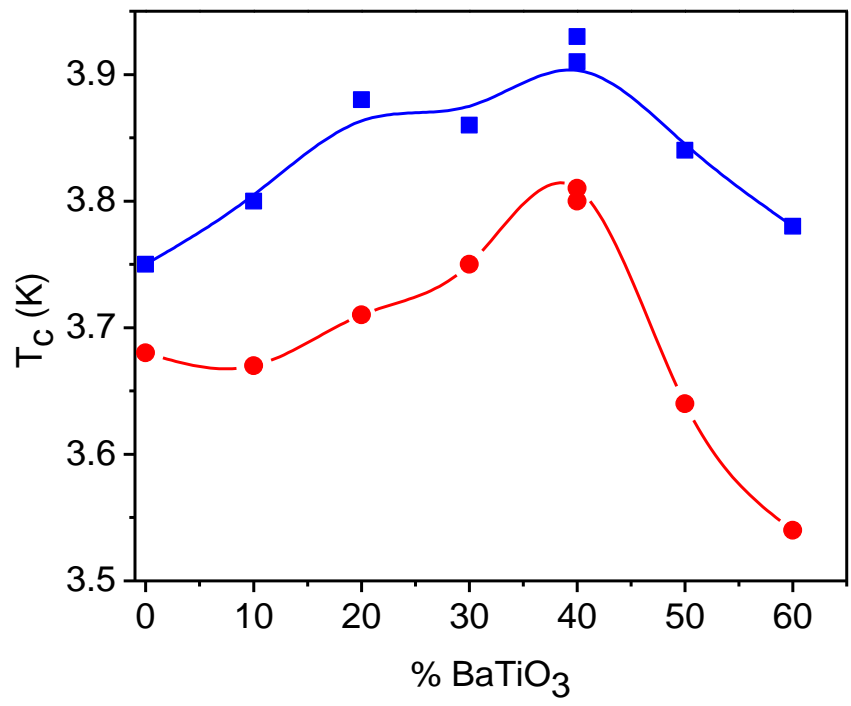

Fig. 5

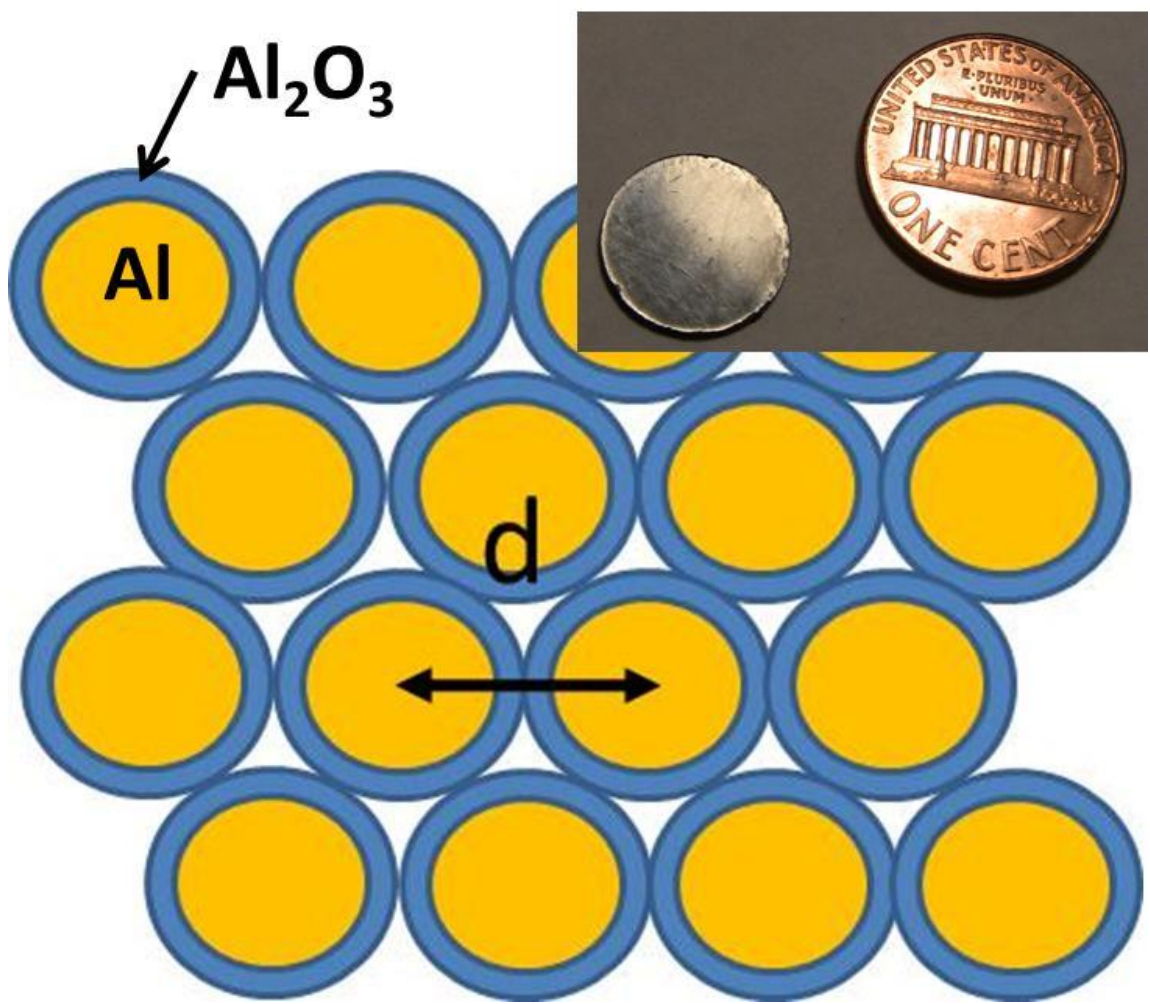

Fig. 6

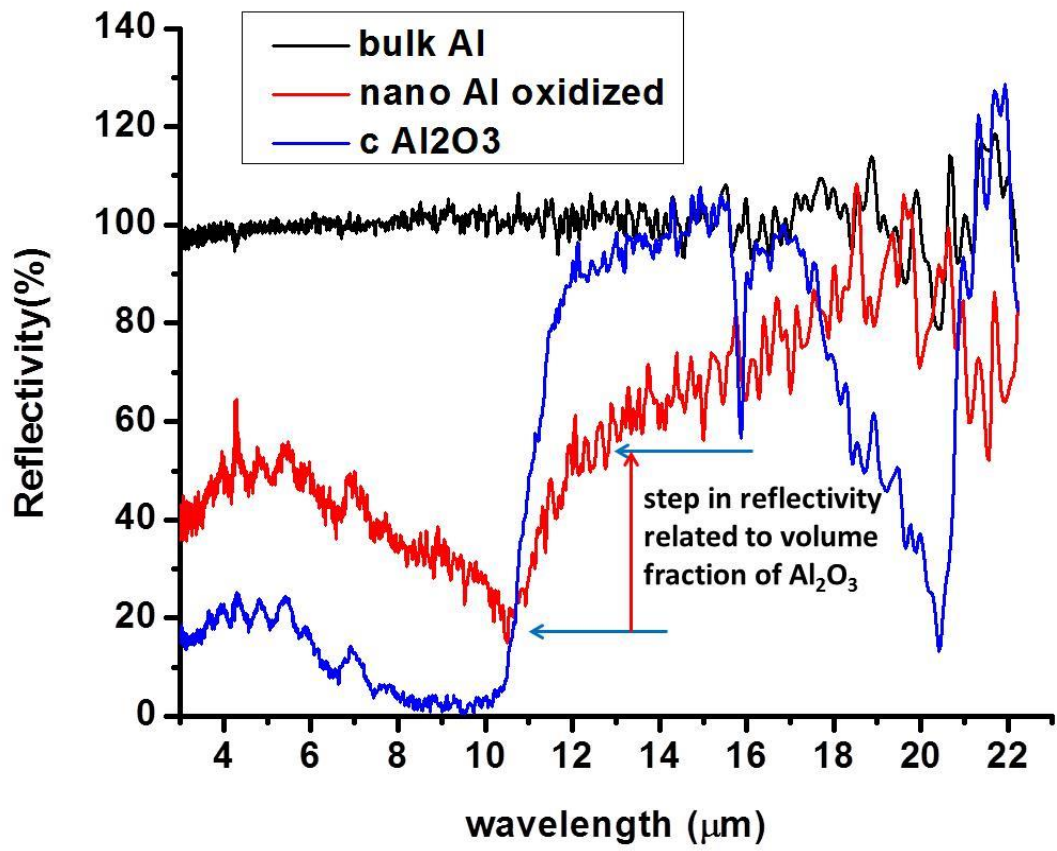

Fig. 7

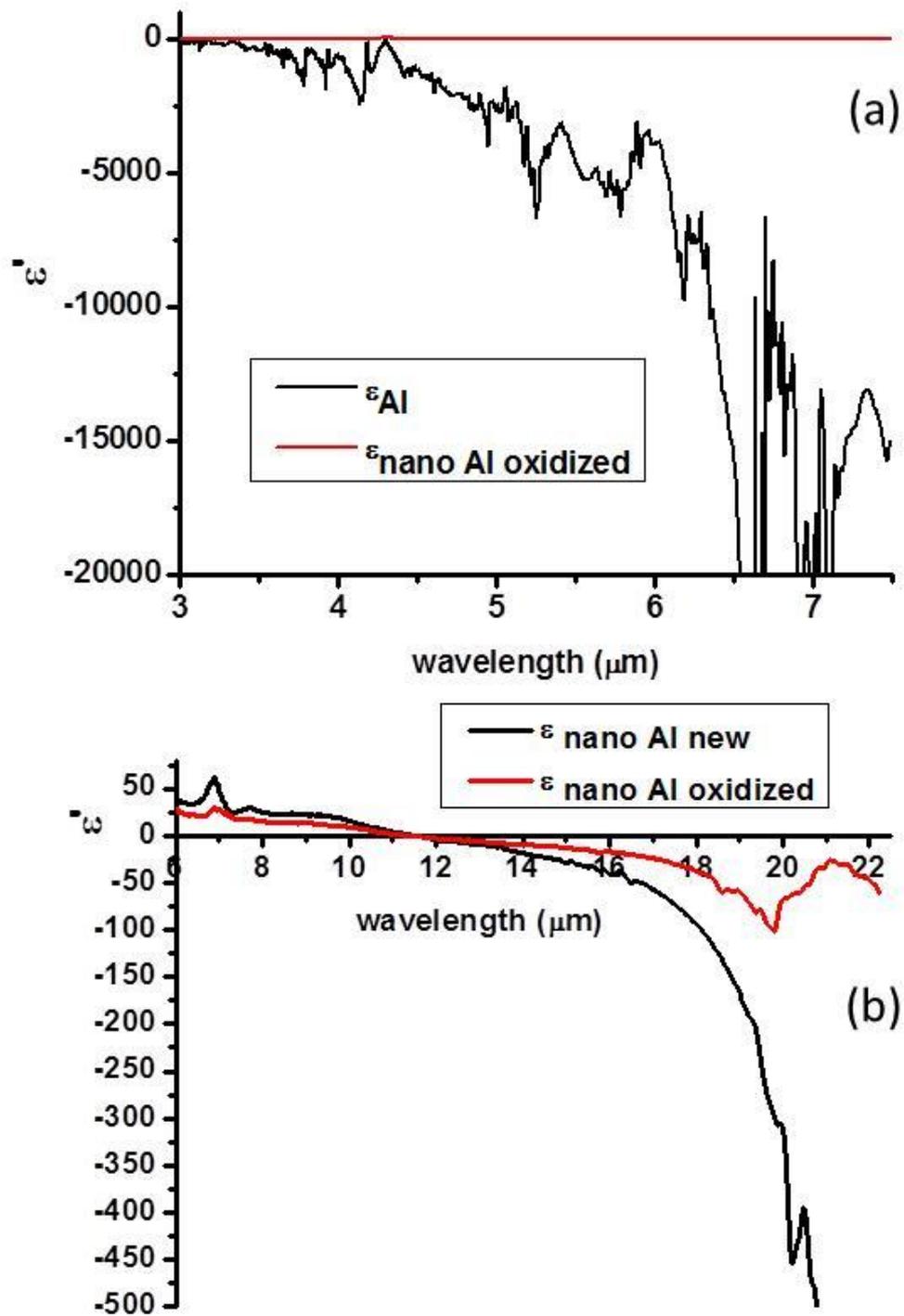

Fig. 8

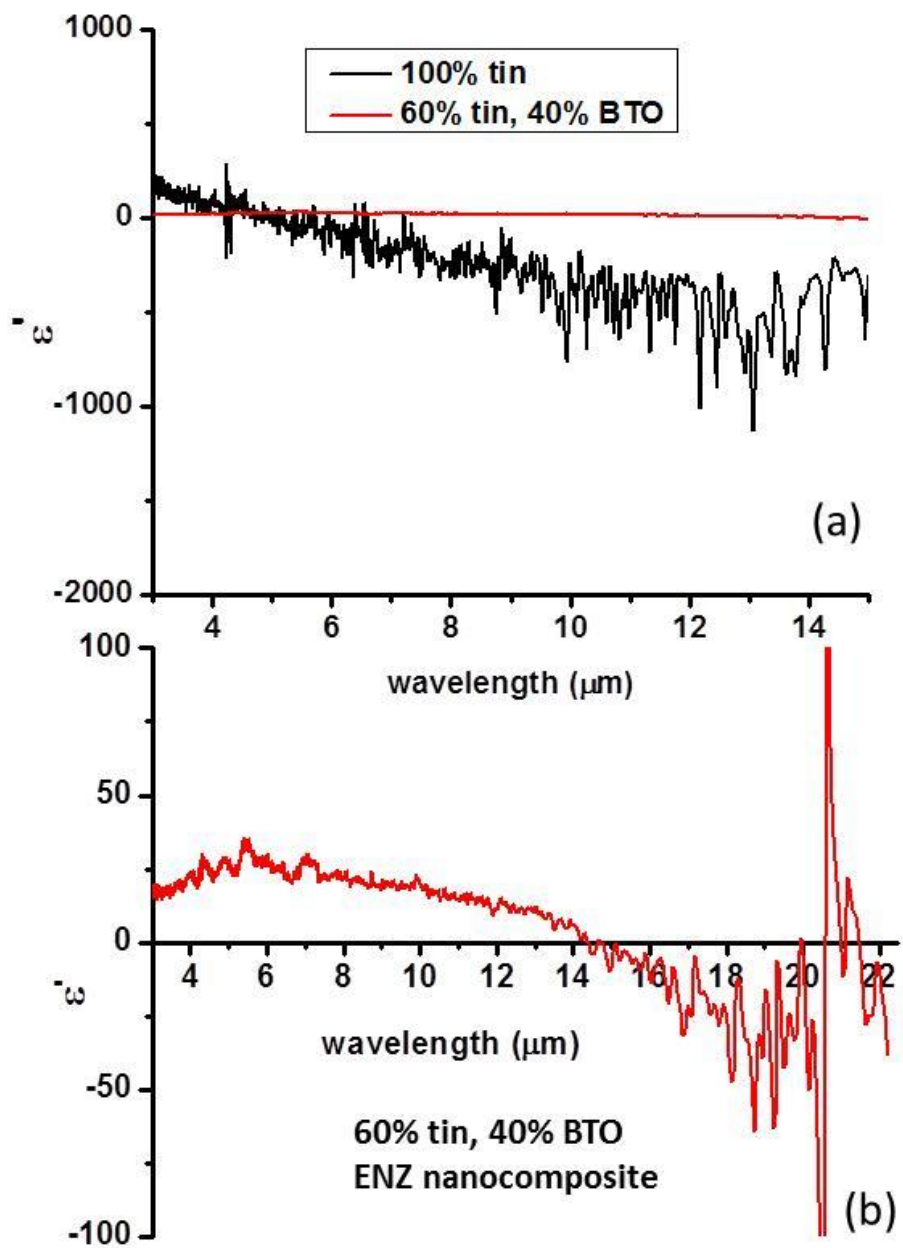

Fig. 9

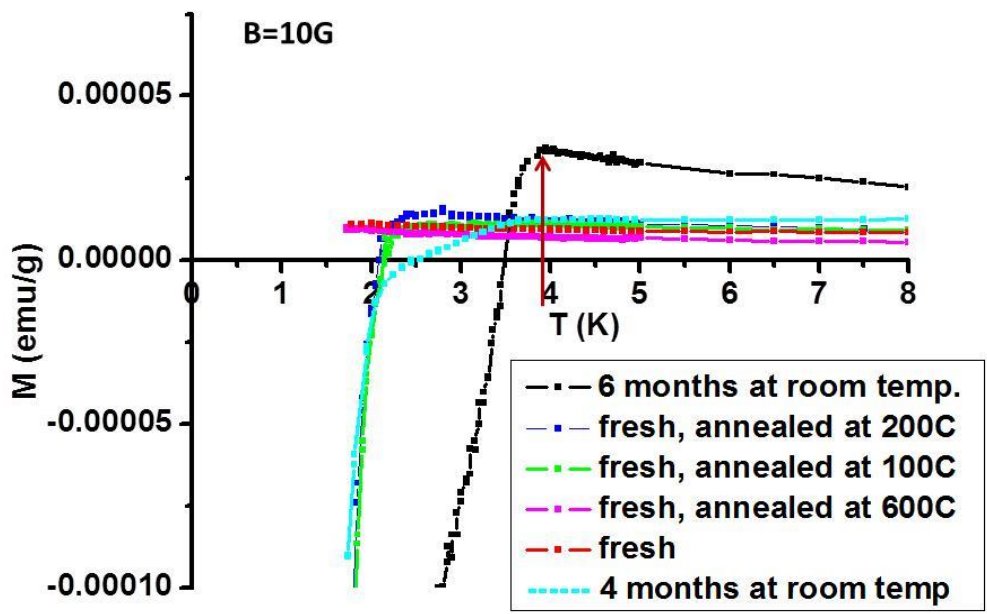

(a)

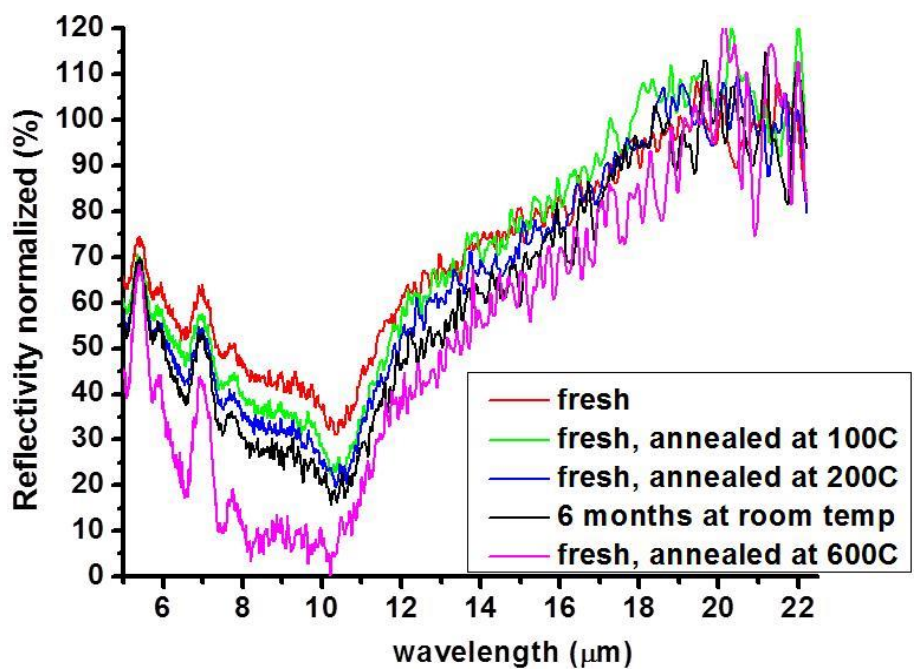

(b)

Fig. 10

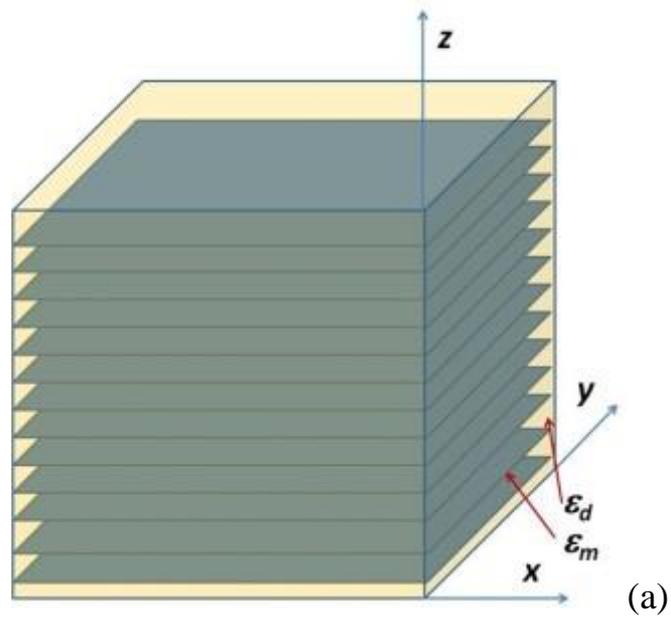

(a)

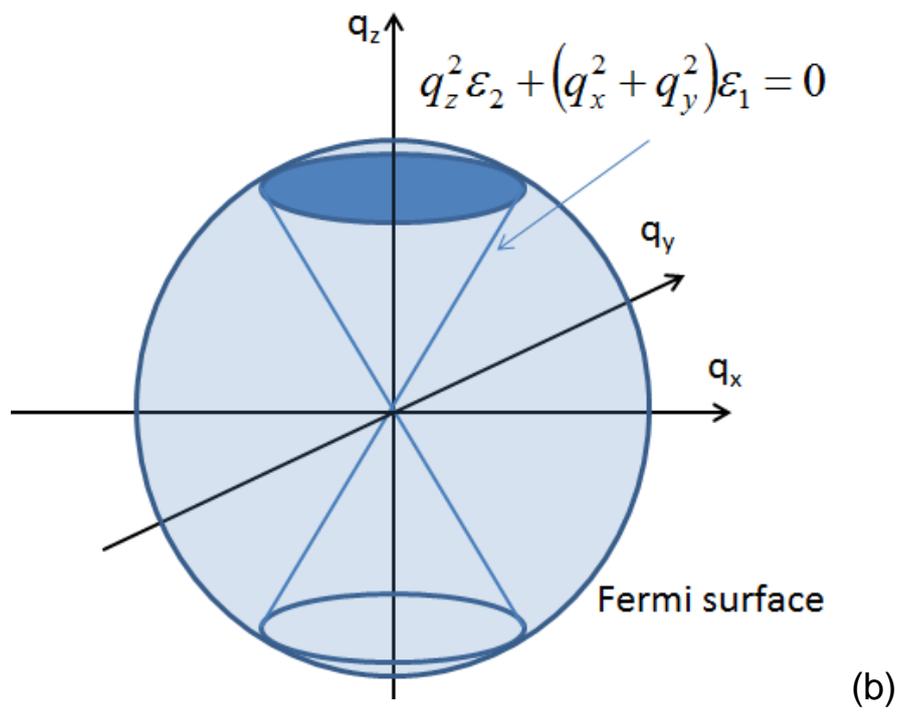

$$q_z^2 \varepsilon_2 + \left(q_x^2 + q_y^2\right)\varepsilon_1 = 0$$

Fermi surface

(b)

Fig. 11

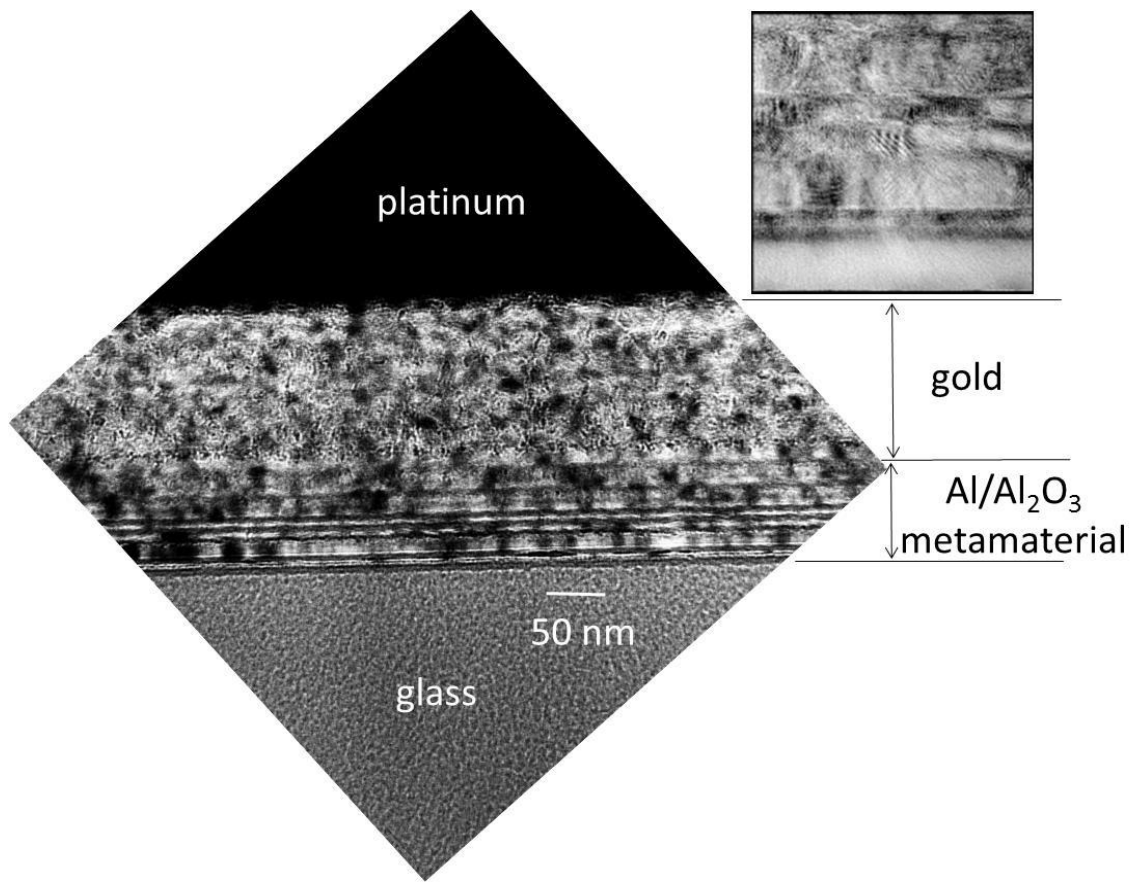

platinum

gold

Al/Al$_2$O$_3$ metamaterial

50 nm

glass

Fig. 12

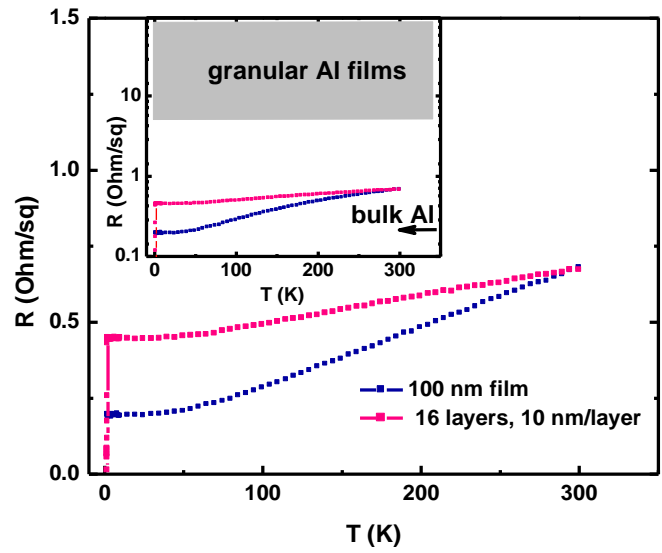

(a)

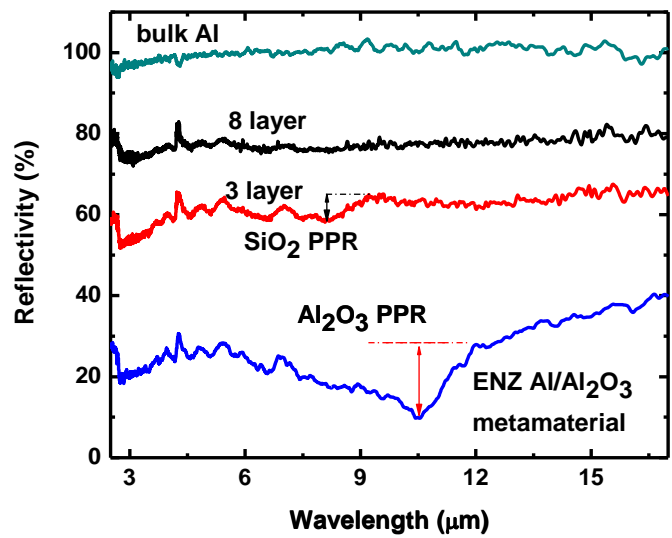

(b)

Fig. 13

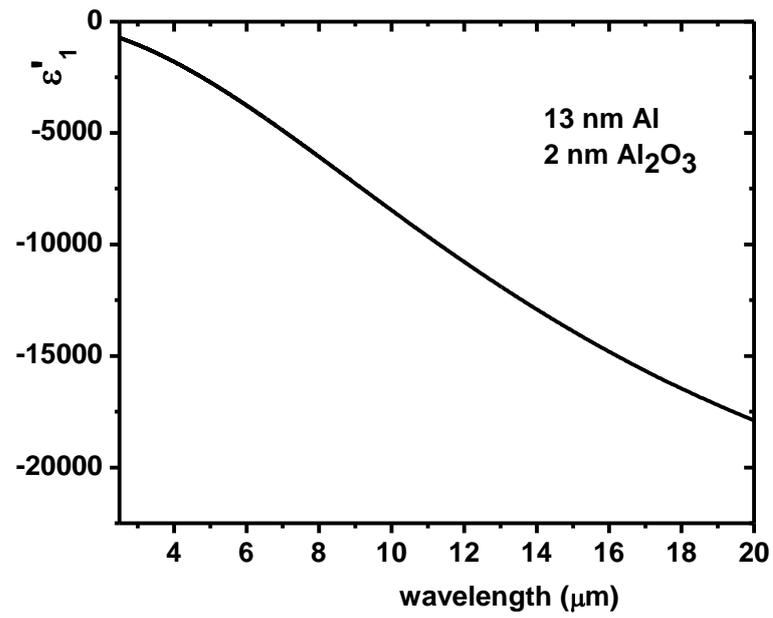

(a)

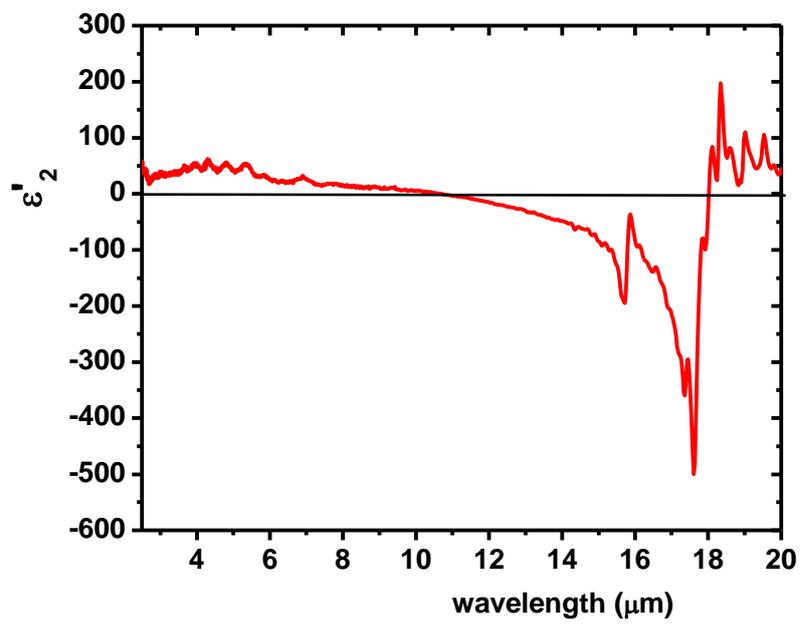

(b)

Fig. 14

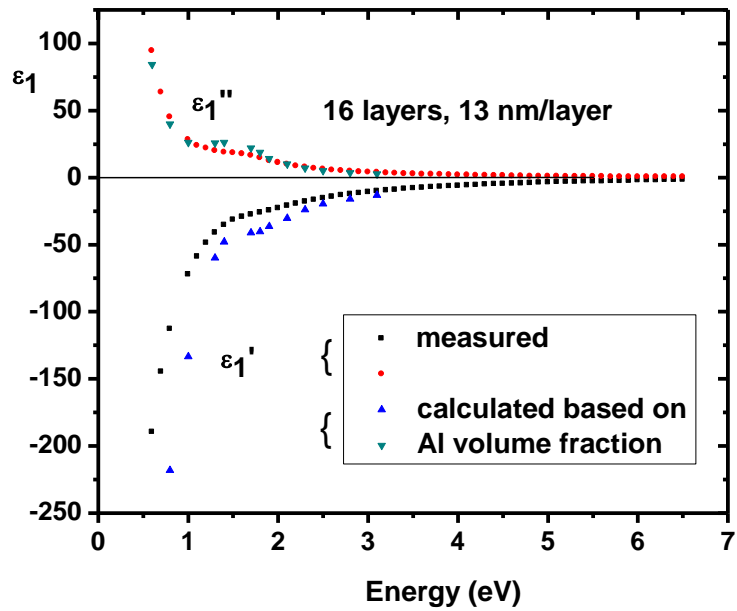

(a)

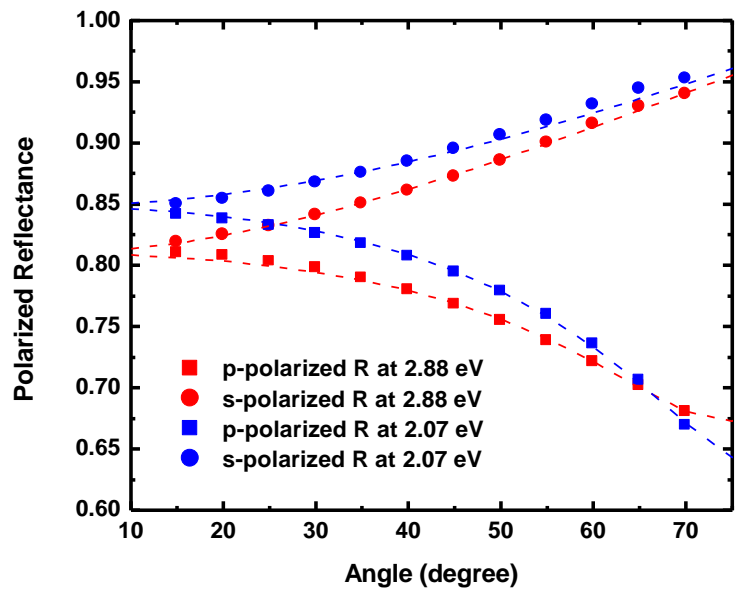

(b)

Fig. 15

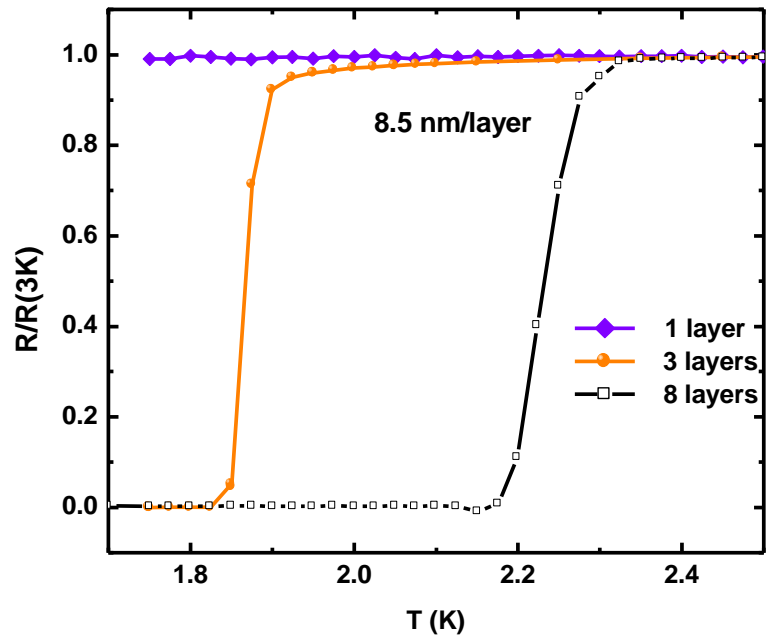

(a)

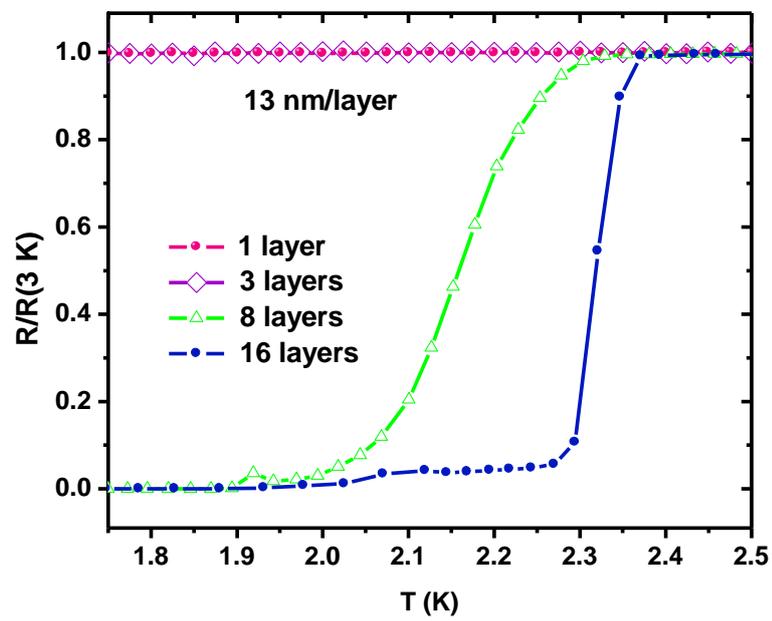

(b)

Fig. 16
.

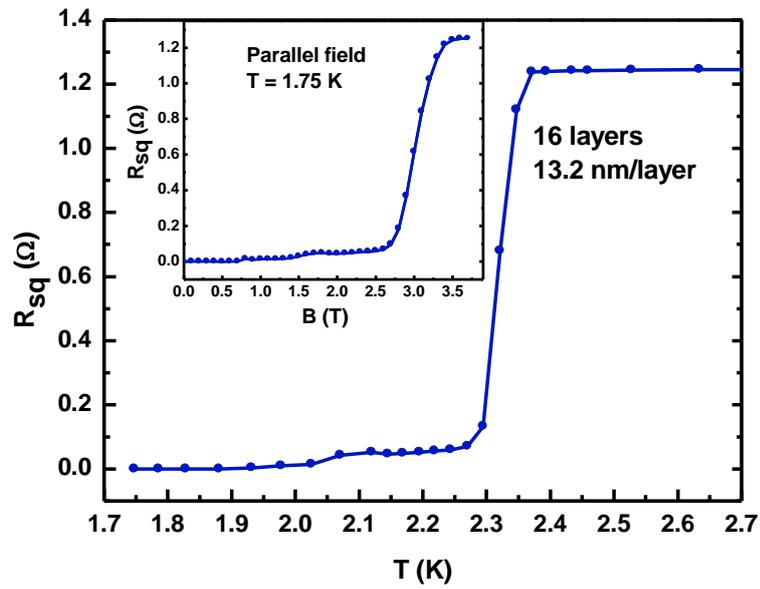

(a)

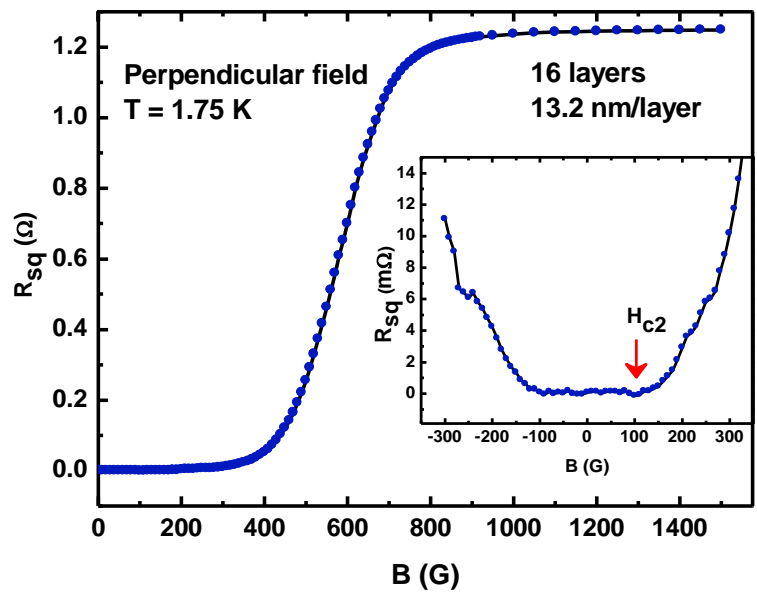

(b)

Fig. 17

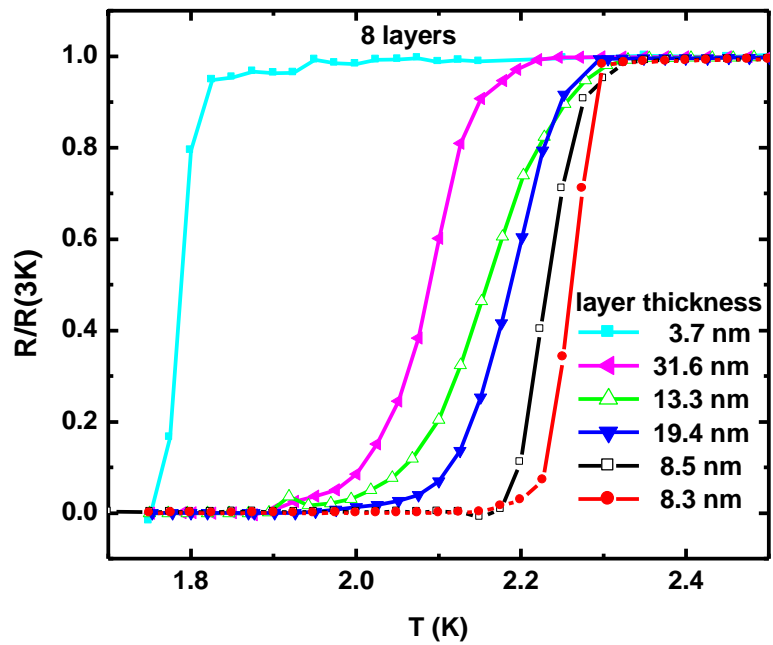

(a)

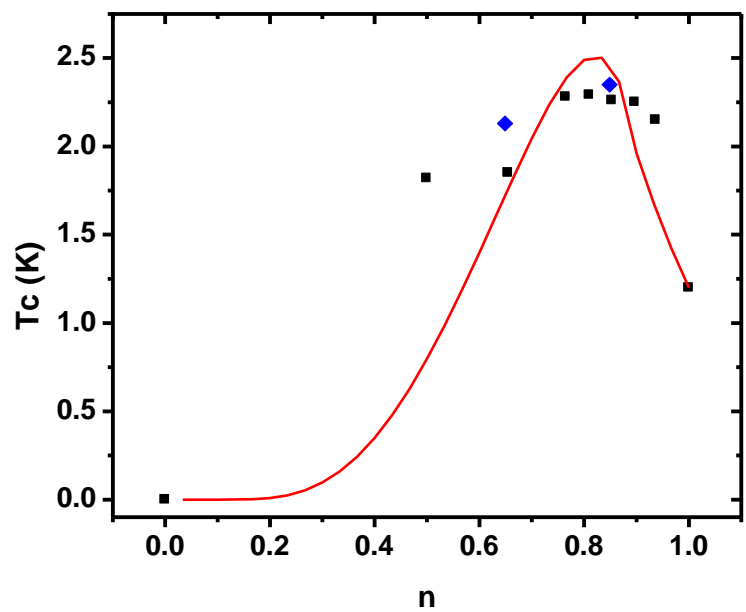

(b)

Fig. 18

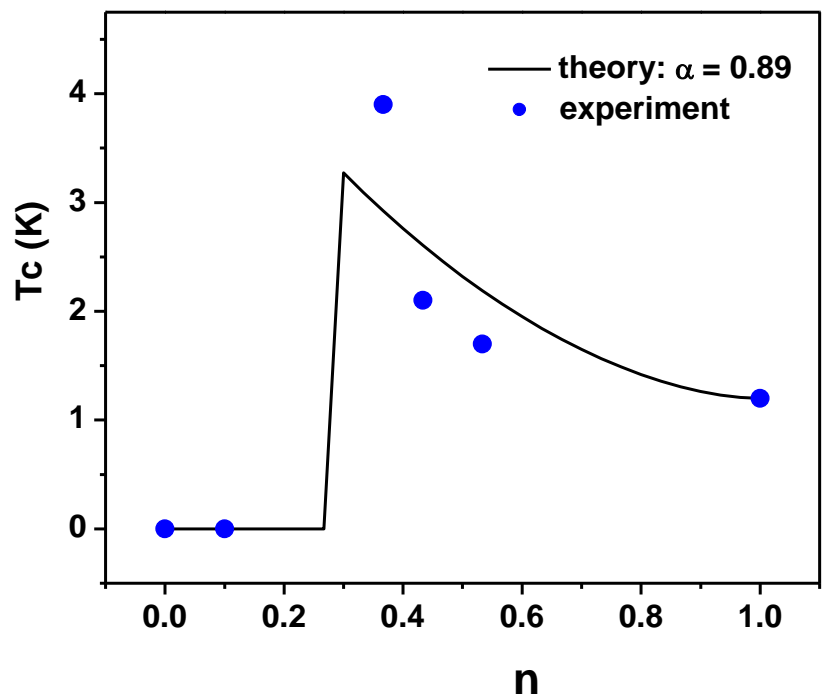

(a)

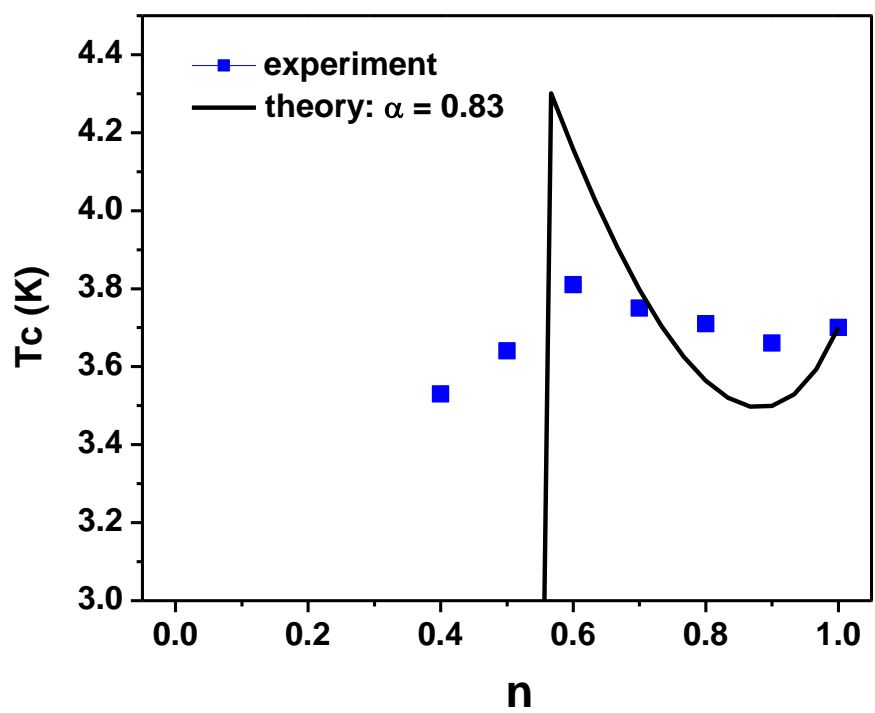

(b)

Fig. 19

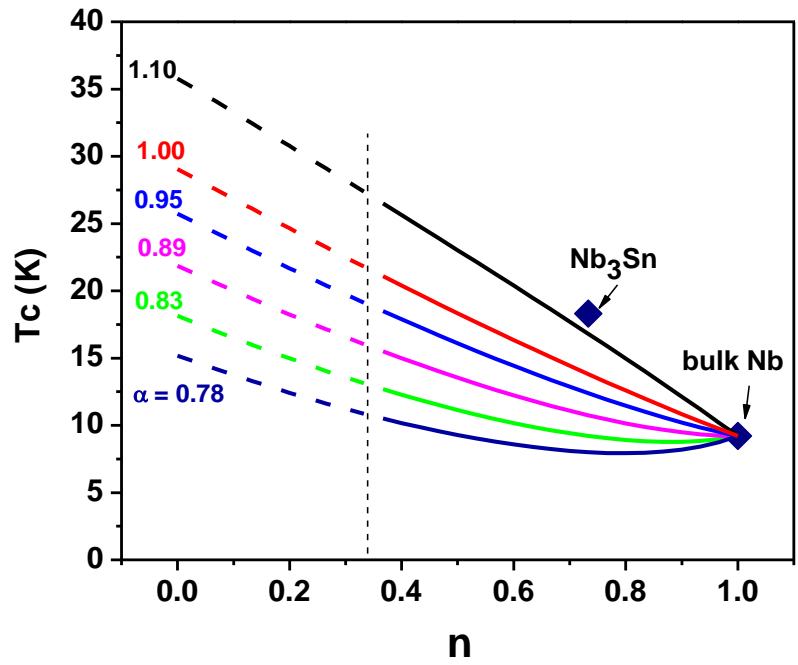

(a)

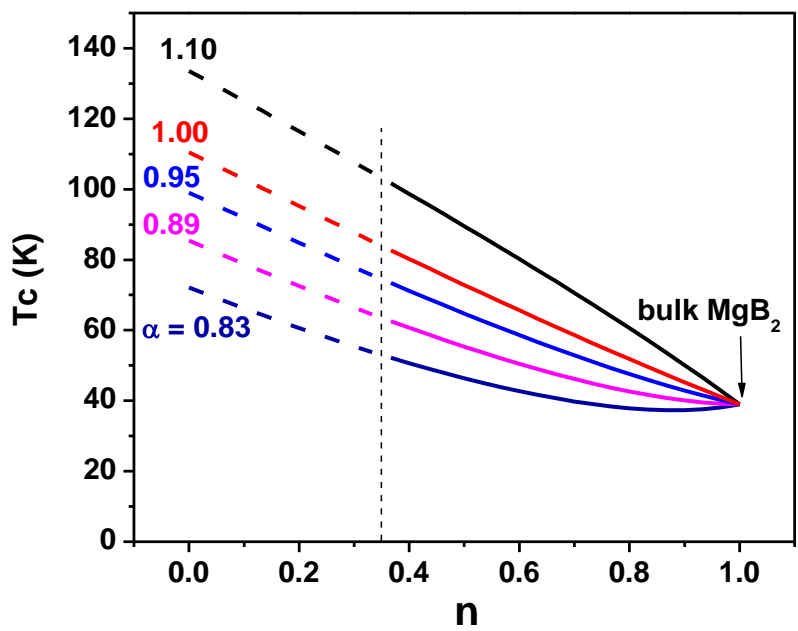

(b)

Fig. 20

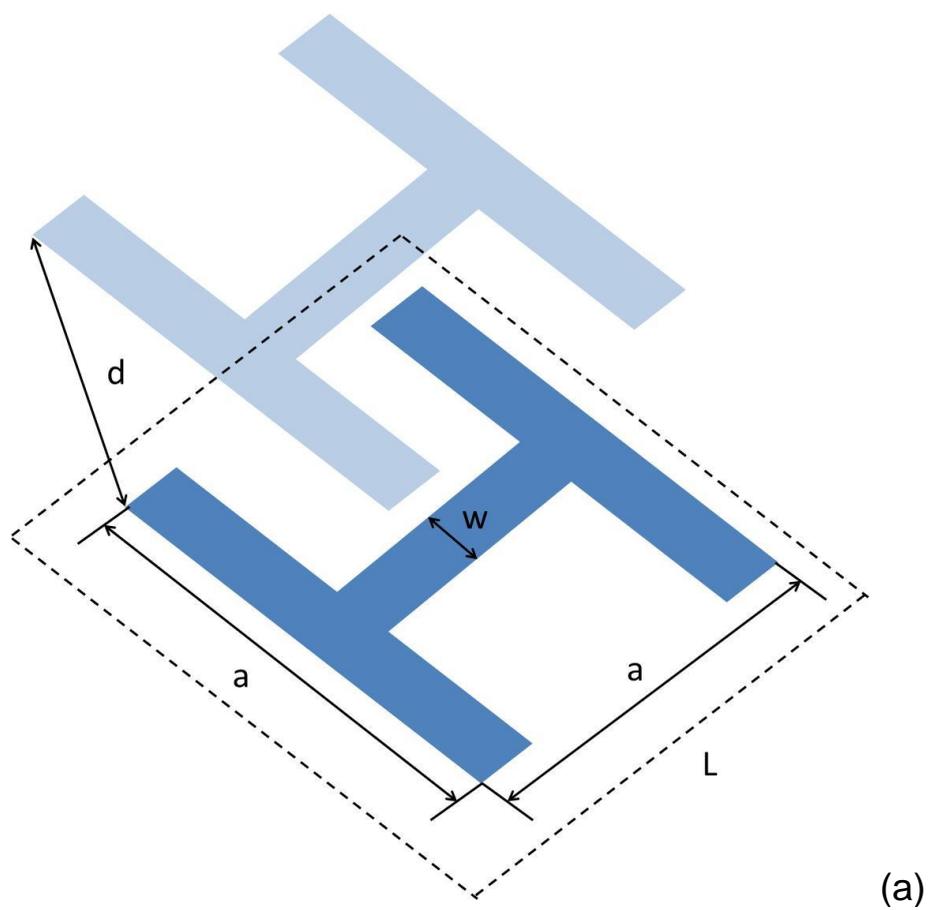

(a)

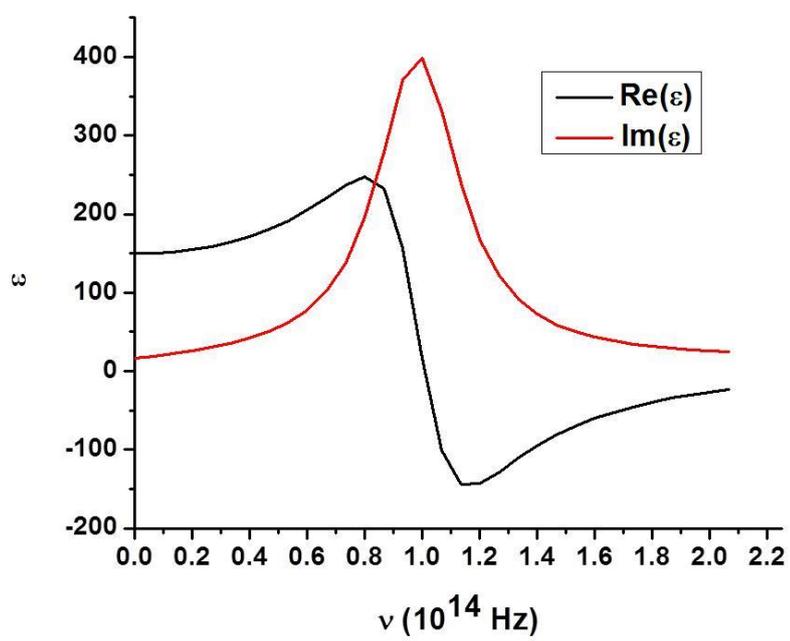

(b)

Fig. 21

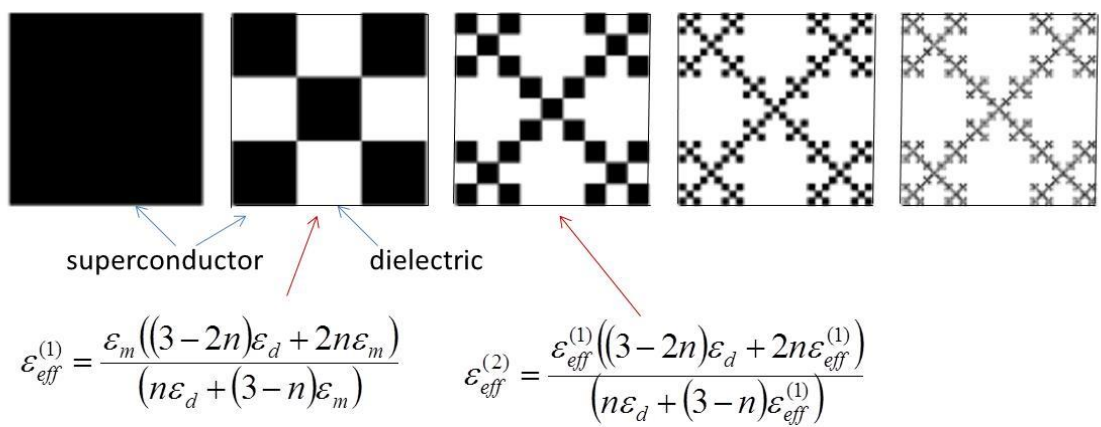

$$\varepsilon_{eff}^{(1)} = \frac{\varepsilon_m\left((3-2n)\varepsilon_d + 2n\varepsilon_m\right)}{\left(n\varepsilon_d + (3-n)\varepsilon_m\right)} \qquad \varepsilon_{eff}^{(2)} = \frac{\varepsilon_{eff}^{(1)}\left((3-2n)\varepsilon_d + 2n\varepsilon_{eff}^{(1)}\right)}{\left(n\varepsilon_d + (3-n)\varepsilon_{eff}^{(1)}\right)}$$

(a)

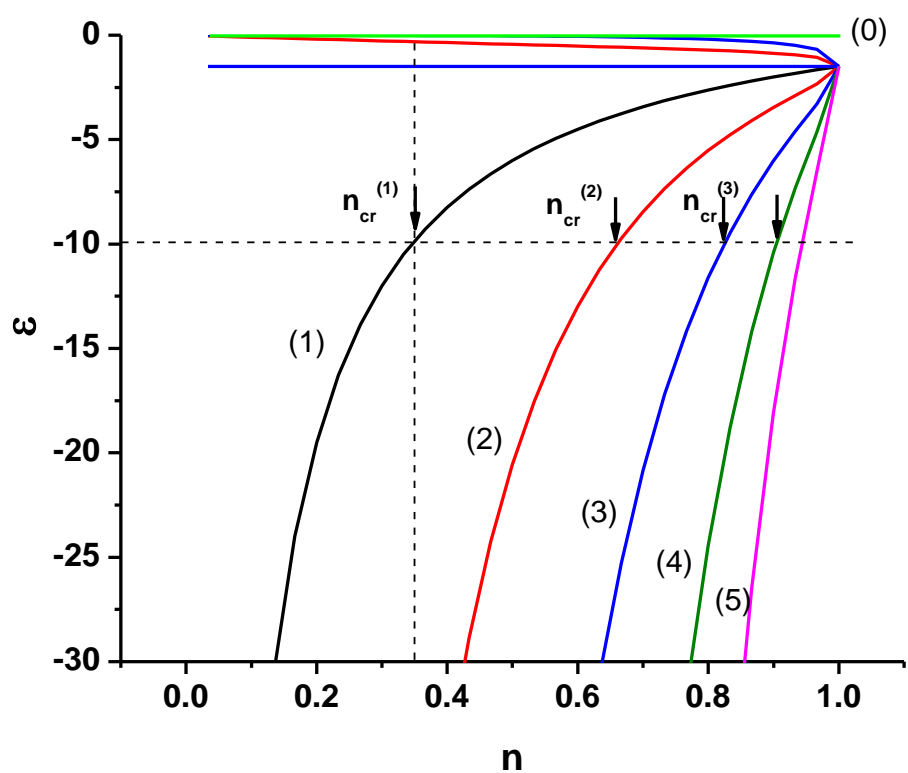

(b)

Fig. 22

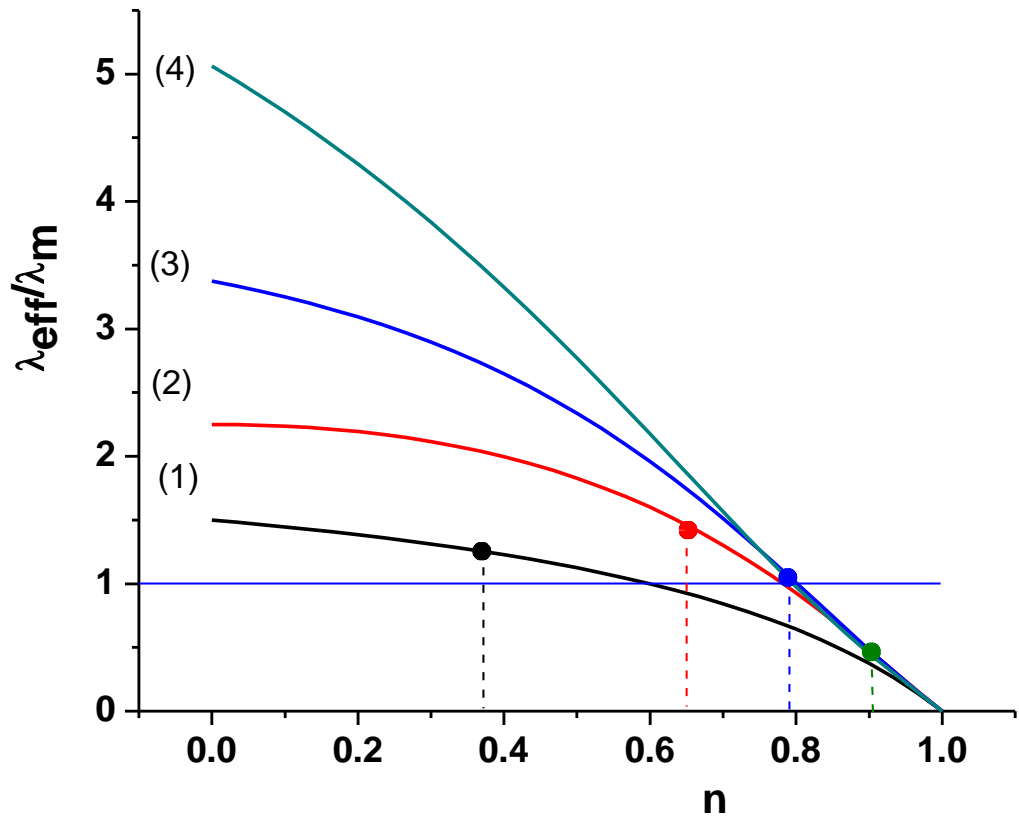

Fig. 23

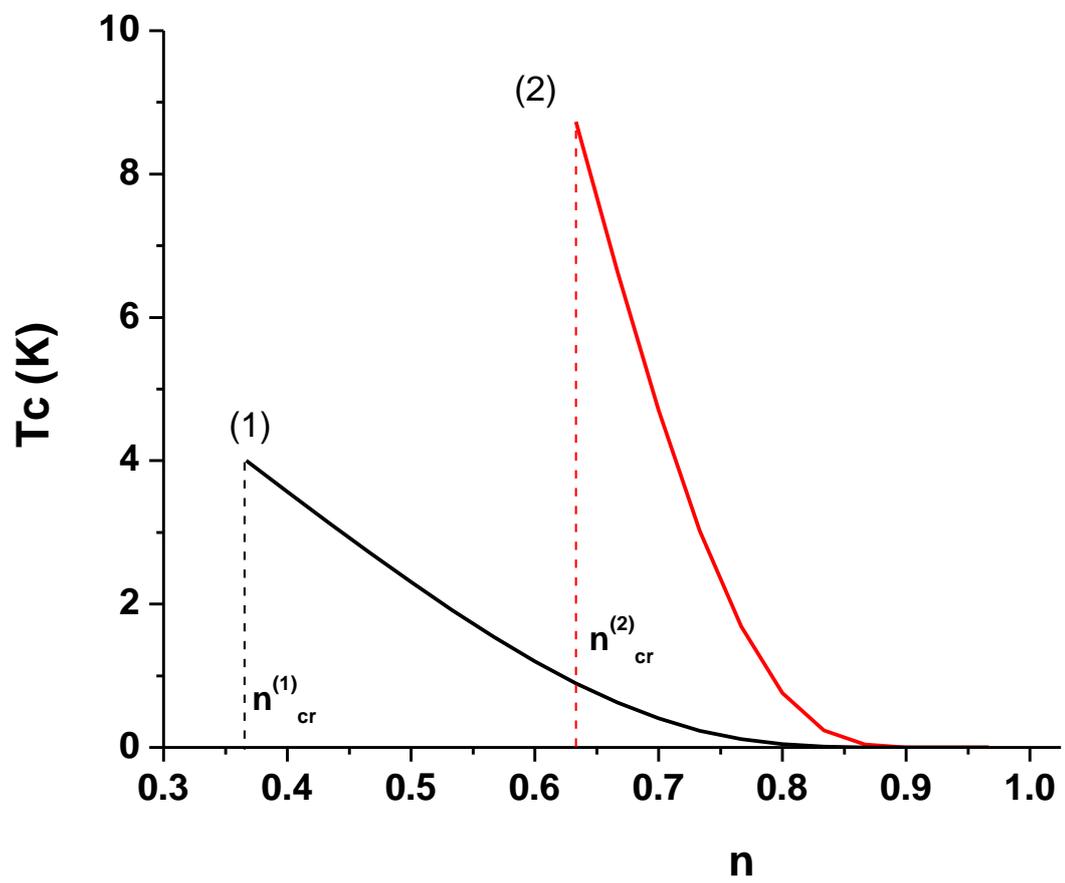

Fig. 24